\documentclass[
aps,
prb,
twocolumn,
floatfix,nofootinbib,
superscriptaddress
]{revtex4-2}

\usepackage{placeins}
\usepackage{multirow}
\usepackage{verbatim}

\usepackage[utf8]{inputenc}
\usepackage{times}
\usepackage[breaklinks]{hyperref}
\usepackage{color}
\usepackage{siunitx}
\usepackage{booktabs}
\usepackage[version=4]{mhchem}

\usepackage{amssymb} 
\usepackage{amsmath}
\usepackage{amsthm}
\usepackage{braket}

\newcommand{\inm}{Institute for Computational and Systems Neuroscience (INM-6),
Forschungszentrum J\"ulich, 52425 J\"ulich, Germany}
\newcommand{\pgi}{Peter Gr\"unberg Institut and Institute for Advanced Simulation (IAS-1),
Forschungszentrum J\"ulich and JARA, 52425 J\"ulich, Germany}

\newcommand{\aachenphys}{Department of Physics, RWTH Aachen University, 52056 Aachen, Germany}
\newcommand{\aachencomp}{Department of Computerscience, RWTH Aachen University, 52056 Aachen, Germany}

\newcommand{\mainz}{Institute of Physics, Johannes Gutenberg University Mainz, 55099 Mainz, Germany}

\renewcommand{\vec}[1]{\boldsymbol{#1}}
 
\newcommand{\myMat}[1]{\mathcal{#1}}

\newcommand{\abs}{\mathrm{\abs}}

\newcommand{\tr}{\mathrm{tr}}

\usepackage{tikz}
\usepackage{xcolor}
\usepackage{graphicx}

\usepackage[capitalise]{cleveref}

\makeatletter
\renewcommand\@biblabel[1]{#1.}
\makeatother

\begin{document}
	
\setcounter{secnumdepth}{2} 


\title{Machine learning inspired  models for Hall effects in non-collinear magnets}

\author{Jonathan Kipp}
\affiliation{\pgi}
\affiliation{\aachenphys}

\author{Fabian R. Lux}
\affiliation{\mainz}

\author{Thorben P\"urling}
\affiliation{\mainz}

\author{Abigail Morrison}

\affiliation{\inm}
\affiliation{\aachencomp}

\author{Stefan Bl\"ugel}

\affiliation{\pgi}

\author{Daniele Pinna}
\affiliation{\pgi}

\author{Yuriy Mokrousov}

\affiliation{\pgi}
\affiliation{\mainz}
\begin{abstract}
	The anomalous Hall effect has been front and center in solid state research and material science for over a century now, and the complex transport phenomena in nontrivial magnetic textures have gained an increasing amount of attention, both in theoretical and experimental studies. However, a clear path forward to capturing the influence of magnetization dynamics on anomalous Hall effect  even in smallest frustrated magnets or spatially extended magnetic textures is still intensively sought after. In this work, we present an expansion of the anomalous Hall tensor into symmetrically invariant objects, encoding the magnetic configuration up to arbitrary power of spin. We show that these symmetric invariants can be utilized in conjunction with advanced regularization techniques in order to build models for the electric transport in magnetic textures which are, on one hand, complete with respect to the point group symmetry of the underlying lattice, and on the other hand, depend on a minimal number of order parameters only. Here, using a four-band tight-binding model on a honeycomb lattice, we demonstrate that the developed method can be used to address the importance and properties of higher-order contributions to transverse transport. The efficiency and breadth enabled  by this method provides an ideal systematic approach to tackle the inherent complexity of response properties of noncollinear magnets, paving the way to the exploration of electric transport in intrinsically frustrated magnets as well as large-scale magnetic textures.
\end{abstract}

\maketitle

\date{\today}

\section{Introduction}
The anomalous Hall effect (AHE) has been an essential measurement tool to study magnetic matter since its discovery over 100 years ago, exhibiting a plethora of mechanisms responsible for transverse electric transport 
\cite{Garello2013_SOT}. The theoretical description of this ubiquitous phenomenon has an equally long history of new revelations, offering an ever modernizing view of electric transport in magnetic materials, with deeper insight into material characteristics gradually accumulating.
	
While extrinsic, scattering driven contributions to transverse resistivity can dominate in certain regimes, it is the contribution rooted in the intrinsic electronic structure that provides a large part of the signal in many cases, especially in strongly spin-orbit coupled materials \cite{Yao2004_FEAHE,CzajaAHE}. Investigations of intrinsic AHE are inseparably linked to the study of the geometry and topology of electronic states, most prominently condensing in Berry-phase effects emerging as a result of various flavors of emergent fields. Although AHE is traditionally associated with Berry phase in reciprocal space, the celebrated topological Hall effect (THE) is related to the Berry phase electrons acquire in magnetic textures with nonzero scalar spin chirality $\xi=\vec{s}_A\cdot(\vec{s}_B\times\vec{s}_C)$ in real space \cite{Bruno2004}. While the emergent field picture has had great success particularly due to its conceptual elegance, experimental evidence suggesting the necessity to expand this approach, especially in materials with strong spin-orbit interaction, is accumulating rapidly~\cite{Lux2018,Lux2020,Yamakage2021,TokuraPRB2021,JubaPRL2021}.	
	
In the past years, the field of AHE in complex magnetic materials has been experiencing a true uprising~\cite{Libor2022AHEAlterMagnet}. The key observation is that a combination of crystal symmetries with time-reversal symmetry breaking by non-collinear magnetic order can result in contributions to the AHE which go beyond those traditionally associated with ferromagnetic magnetization or simplified real-space emergent field pictures. One of  suggested guiding principles in keeping a systematic overarching counting of possible contributions to the AHE in a system which undergoes drastic changes in its magnetic configuration lies in a symmetry expansion of the anomalous Hall conductivity in terms of products of spins on different atomic sites of arbitrary power~\cite{Kipp2021chiralComPhys}. This approach has been applied in the past to the case of a bipartite lattice, demonstrating that it is possible to conceptually separate the AHE in systems with two spins into chiral and crystal contributions~\cite{Kipp2021_SpinSpirals,Lux2022_AHEMnBiTe}, which collect different powers of the spin expansion. Demonstrating the intricate physics that lies behind this decomposition, it was in particular shown that the corresponding chiral Hall effect (CHE) intertwines together the Berry phases of Bloch electrons in real and reciprocal spaces \cite{Lux2020}. While being very rigorous and potentially very promising in uncovering the physics of higher-order contributions to the AHE, applying this method to systems containing more spins ideally requires an automated approach. 

Generally, the research in the area of AHE can benefit greatly from a deep understanding of the implications that crystal symmetry has on the geometry and topology of Bloch electrons, which can be modified excessively by tuning parameters in the enormous phase space of the magnetic configurations. While this phase space offers intriguing opportunities for exploration~\cite{Malottki2017_EnhancedSS,Bessarab2015_MagTrans}, the numerical complexity grows exponentially with the number of magnetic moments in the system.
Increased computational cost for large system sizes or complex combinatorics are problems many disciplines suffer from as a whole. Consequently, natural sciences, and especially computational solid state physics, are witnessing a paradigm shift to find a way around increasingly costly simulations. Machine learning techniques have been gaining more and more attention over the past 10 years and have found application particularly in the search for pretrained electron potentials for accelerated density functional theory (DFT) calculations \cite{Czany2010_GaussianPotentials}, computational materials design \cite{Gigli2022_Batio3,Hongbin2022_MLHighEntropyalloys,Hilgers2023_BatchLearning,Hilgers2023_HeuslersCritTemp}, rapid exploration of large parameter spaces \cite{MarquardtRapidExplo2021}, or interpretation of extensive experimental databases \cite{Kim2022_MLXray}. However, comparable studies focusing on magnetic degrees of freedom are few and far between \cite{Li2020_EffectiveSpinHamil}. 

The machine learning techniques have matured so far that feature selection, feature extraction and model regularization can distill relevant features from high-dimensional input both effectively and reliably, and therefore aid algorithms which are in principle unaware of any physical intuition in finding models which we would accept as physical. 
In essence, using the right tools out of a huge, readily available toolbox, it is possible to fit models depending on just the relevant fraction of a large number of features, thereby reducing a phase space which might be too large to conquer to manageable size. Consequently, the main objective of this work is to train minimal, regularized models predicting the electric transport properties of complex magnets in terms of suitable descriptors with machine learning techniques.

At the heart of this modelling framework an expansion in symmetric invariants is employed as a basis for the descriptor space. This kind of expansion is a powerful ingredient in the sense that it confines the shape of the terms being used, but at the same time provides us with all possible spin interaction terms respecting the crystal symmetry. It is therefore a valuable alternative to spin modelling techniques, where product terms are constructed ``by hand``  \cite{Li2020_EffectiveSpinHamil}, because it is complete with respect to the point group symmetry.
In magnetism, such atomistic spin models traditionally offer great utility for the study of magnetic properties in realistic materials, and recently, efforts have been made to optimize these spin Hamiltonians with the help of machine learning techniques \cite{Li_2023_MultiFerro}. However, the issues surrounding the role of the magnetic reference state used for the calculations, performed in order to extract the interactions parametrizing the spin Hamiltonian have been subject to debate~\cite{DosSantos2022_ChiralMagInts,Katsnelson2023_MagnIntReview}. Accordingly, expansion techniques which do not refer to a fixed reference state, like the one presented in this work, might offer a new perspective on one of the staple methods in magnetism research.

Furthermore, the expansion method is not bound to a specific limit, neither microscopically canted spins nor mesoscopically extended magnetic textures, and might therefore be employed to derive a description which transitions smoothly between the theories for the two limits.
Ultimately, the main goal of this project is to obtain linear models with reasonable generalization statistics on unseen data for calculations of the AHE in complex canted magnets with strong spin-orbit interaction, using a minimal number of descriptors constructed from the symmetric invariant expansion. From these models we aim to extract, first,  the already familiar contributions to the AHE, as well as further, previously unknown terms in higher orders of the expansion. Second, we will test whether these models can be reliably and efficiently employed for extrapolation of the AHE in the magnetic phase space. Third, by varying the Fermi energy, we will examine whether the model construction is consistent enough to enable the exploration of the AHE in the combined phase space of spin moments and that of Bloch states with intricate reciprocal space geometry.

Mapping the fitted model coefficients as functions of the tight-binding model parameters as well as the system size will give us insight into the behavior of the different contributions to the AHE in different regimes. 
We classify the many higher order terms in low dimensional, microscopically canted magnets with respect to vector chirality $\vec{\xi}=\vec{s}_A\times\vec{s}_B$ in order to derive a recipe for predicting the magnetotransport properties in large scale magnetic textures, such as spin spirals, skyrmions, or multi-q states. While we focus here on the chirality-even contributions to the AHE, which correspond e.g. to the so-called crystal Hall effect in case of antiferromagnets~\cite{Smejkal_crystal_2019,Kipp2021chiralComPhys}, one of our goals lies in providing an ability to identify the differences in transport signatures between different textures and associate these differences with specific $\vec{\xi}$-odd or -even terms in the model. This will contribute to identification and distinction of magnetic textures by measurement of electric transport, which might play a key role in development of novel computing architectures.

In our work we bring together aspects form multiple fields, namely the calculation of the AHE from an electronic model, the expansion of a tensor in symmetric invariants, in this specific case the anomalous Hall conductivity tensor, and the modelling of the tensor based on invariant expansion utilizing machine learning methods. Therefore, we begin by introducing the electronic model and its symmetries in \cref{sec:Model_and_Syms} as well as the linear response method used in the calculation of the anomalous Hall effect in \cref{sec:Kubo}. We continue with illustrating the representation-based expansion machinery in \cref{sec:Expansion}, before explaining the details of the model selection pipeline in \cref{sec:Pipeline}. Finally, we present and analyze fitting results in \cref{sec:FitResults}. We comment on the feature scaling with respect to parameters of the tight-binding model and offer a concluding discussion in \cref{sec:Conclusion}.

\section{Method}

In our work we draw inspiration from three different perspectives, which define the three levels of our approach: on the surface, this is a machine learning approach to fitting sparse, linear models. The second level reveals, that input features to the learning algorithm are supplied by an expansion of the target in orders of spin, firmly tied to the underlying lattice symmetry. The base level of this framework provides the target to the algorithm, by electronic structure calculations on a discrete, magnetic lattice. Defining a working process to successfully combine these three perspectives in a modelling pipeline is at the core of our manuscript. 
In order to clarify the multifaceted approach presented here, we define the working parts of the pipeline first, before illustrating the methods in obtaining training data later in \cref{sec:PipelineResults}.
\subsection{Electronic model and its symmetries}\label{sec:Model_and_Syms}
In this work, we aim at predicting the electric transport properties of electrons on a bipartite honeycomb lattice of magnetic spins. To model the electronic structure, we employ  an  effective two-dimensional lattice  tight-binding (TB) Hamiltonian (in the $xy$-plane) which reads:
\begin{equation}
\begin{split}
	H = -t \sum\limits_{\langle ij \rangle\alpha}  c_{i\alpha}^\dagger c_{j\alpha}^{\phantom{\dagger}} &+ i \alpha_{\rm R}\sum\limits_{\langle ij \rangle\alpha \beta}  \hat{\mathbf e}_z \cdot (\boldsymbol{\sigma} \times {\mathbf d}_{ij})_{\alpha\beta}\, c_{i\alpha}^\dagger c_{j\beta}^{\phantom{\dagger}}\\
	&+ \lambda_{\rm ex} \sum_{i\alpha \beta} (\hat{\mathbf s}_i\cdot \boldsymbol{\sigma})_{\alpha\beta}\, c_{i\alpha}^\dagger c_{i\beta}^{\phantom{\dagger}},
\end{split}
\label{eq:model}
\end{equation}
where $c_{i\alpha}^\dagger$ ($c_{i\alpha}^{\vphantom{\dagger}}$) denotes the creation (annihilation) of an electron with spin $\alpha$ at site $i$, $\langle ...\rangle$ restricts the sums to nearest neighbors, the unit vector $\mathbf d_{ij}$ points from $j$ to $i$, and $\boldsymbol{\sigma}$ stands for the vector of Pauli matrices. Besides the hopping with amplitude $t$,  \cref{eq:model} contains the Rashba spin-orbit coupling of strength $\alpha_\text{R}$ originating for example in the surface potential gradient perpendicular to the plane (i.e.\,along $\hat{z}$). The remaining term in \cref{eq:model} is the local exchange term with $\lambda_{\rm ex}$ characterizing the strength of exchange splitting and $\hat{\mathbf{s}}_i$ stands for the direction of spin on site $i$.  Here, we work with the following parameters of the model: $t=1.0$\,eV, $\alpha_{\rm R}=0.4$\,eV, and $\lambda_{\rm ex}=1.4$\,eV.

\begin{figure}[t!]
\includegraphics[width=0.7\hsize]{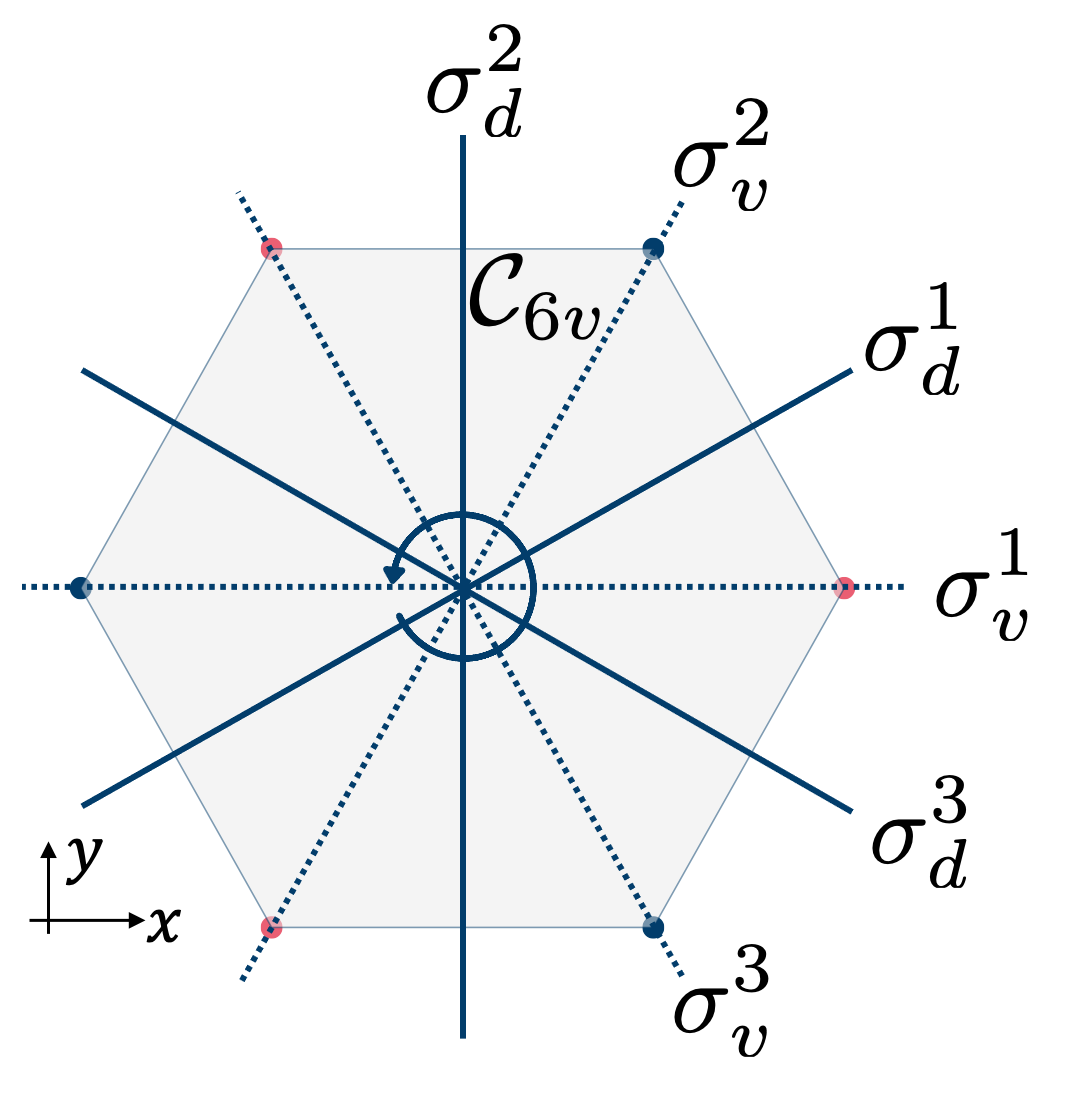}
\caption{{\bf Illustration of symmetry operations of the honeycomb lattice.} The symmetry operations leaving the bipartite honeycomb lattice (atom types in red and blue) invariant can be illustrated on the plane containing the lattice, here the $xy$-plane. The point group of the honeycomb lattice, $C_{6v}$, shows three classes of rotation around the $z$-axis, namely two-, three- and sixfold rotations $C_2(z)$, $C_3(z)$ and $C_6(z)$. Furthermore, there are two classes of mirror operations, $\sigma_v$ and $\sigma_d$, indicated here by the mirror plane going through the corners of the hexagon ($\sigma_v$) or through the midpoint of the hexagons edges ($\sigma_d$).}
\label{fig:Hex_SymOps}
\end{figure}

The analysis presented in this work also heavily relies on the crystallographic point group symmetry of the bipartite, nonmagnetic, honeycomb lattice, which is illustrated in \cref{fig:Hex_SymOps}. We can identify five nontrivial families of symmetry, or conjugacy classes, in the lattice, comprising the point group of $C_{6v}$, see \cref{tab:character_table}: on one hand, the lattice shows three classes of rotational symmetry, namely the six-, three-, and twofold rotation around an axis perpendicular to the lattice plane, which is commonly chosen to be the $z$-axis, with the lattice lying in the $xy$-plane. On the other hand, the nonmagnetic lattice is invariant under two classes of mirror operations, with the mirror plane being aligned with either the corners of the hexagon or the midpoints of the hexagons edges. 

\subsection{Expansion in terms of symmetric invariants}\label{sec:Expansion}
\begin{table}[t!]
\centering
\caption{{\bf Character table of $C_{6v}$.} 
	Shown are the characters of each irreducible representation for each conjugacy class, alongside linear and quadratic basis functions which generate the respective representations. A tensor inducing a particular representation of the point group transforms accordingly to the characters of this representation. This allows us to easily augment the dataset by applying the symmetry operations to the spin configurations and multiply the target value by the according character value of the representation. In the case of the anomalous Hall conductivity (AHC) tensor, which belongs to the representation $A_2$, the tensor changes sign under the mirror operations with the mirror planes cutting through the edges and the middle of the sides of the hexagon, $\sigma_v$ and $\sigma_d$.}
\label{tab:character_table}\vspace{0.2cm}
\begin{tabular}{c|cccccc|c|c}
	$C_{6v}$ & $E$ & 2$C_6\text{(z)}$ & $2C_3\text{(z)}$ & $C_2\text{(z)}$ & 3$\sigma _v$ & 3$\sigma _d$ & linear & quadratic \\ \toprule
	$A_1$ & 1 & 1 & 1 & 1 & 1 & 1   &  $z$ & $x^2+y^2, z^2$ \\
	$A_2$ & 1 & 1 & 1 & 1 & -1 & -1& $n_z$ & \\
	$B_1$ & 1 & -1 & 1 & -1 & 1 & -1 & &\\
	$B_2$ & 1 & -1 & 1 & -1 & -1 & 1& & \\
	$E_1$ & 2 & 1 & -1 & -2 & 0 & 0 &$(x,y)$ & $(xz, yz)$\\
	$E_2$ & 2 & -1 & -1 & 2 & 0 & 0&  & $(x^2 -y^2, xy)$\\ \bottomrule
\end{tabular}
\end{table}

Given an observable of interest, which can be of tensorial nature,  we have to invoke the representation of corresponding tensor in terms of the crystallographic point group and then expand it in the symmetric invariants of this representation, which correspond to the eigenvalues of the invariant subspaces in the representation. Below, we briefly present the basics of representation theory that we make use of in our study~ \cite{SC_Dongwook_2022}.

We call a representation a map $\rho:G\to GL(n,\mathbb{C})$, where $GL(n,\mathbb{C})$ is the general linear group of $n\times n$ matrices with complex elements. $n$ is in this case called the dimension of the representation of $\rho$. Every representation has to be a group homomorphism, meaning that $\rho(g_1g_2)=\rho(g_1)\rho(g_2)$ for all $g_1,g_2\in G$. We speak of two representations $\rho$ and $\rho'$ as equivalent, if there exists a matrix $A$ such that $A\rho(g)A^{-1}=\rho'(g)$, for all $g\in G$. All representations of the same crystallographic point group have an equivalent unitary representation, fulfilling $\rho(g^{-1})=\rho(g)^{\dagger}$, where $\dagger$ denotes complex conjugate transposition.
The character of a representation $\rho$ is defined as the matrix trace $\chi_{\rho}(g)=\tr{\rho(g)}$. The character map is constant under all conjugacy classes, because the trace is cyclic, $\chi(h^{-1}gh)=\chi(g)$, for all $g,h\in G$. For any two representations $\rho_1$ and $\rho_2$ we find
\begin{align}
\chi_{\rho_1\otimes\rho_2}&=\chi_{\rho_1}(g)\chi_{\rho_2}(g)\\
\chi_{\rho_1\oplus\rho_2}&=\chi_{\rho_1}(g) + \chi_{\rho_2}(g),
\end{align}
where $\otimes$ denotes the Kronecker or tensor product and $\oplus$ is the direct sum of matrices. We call a representation reducible if it is equivalent to a direct sum of representations, and irreducible if it is not. The representations $\rho$ of a crystallographic point group can be decomposed into a direct sum of irreducible representations $\Gamma_i$ as follows:
\begin{align}
\rho = \bigoplus_{i=1}^{n_c} \lambda_i\Gamma_i,
\end{align}
where $n_c$ is the number of conjugacy classes of the group. The multiplicity coefficients can be extracted as $\lambda_i=\langle\Gamma_i|\rho\rangle$, since the Schur orthogonality theorem states that
\begin{align}
\langle\Gamma_i|\Gamma_j\rangle = \frac{1}{|G|}\sum_{g\in G}\chi_{\Gamma_i}(g)\chi^{*}_{\Gamma_j}(g)=\delta_{i,j},
\end{align}
with the Kronecker delta $\delta_{i,j}$. One specific representation we are interested in is the magnetic representation, i.e. the product of site and spin representation:
\begin{align}
\Gamma_{\text{mag}}=\Gamma_{\text{site}} \otimes \Gamma_{\text{spin}}.
\end{align}
This representation has the dimension $d=3\cdot n_s$, with $n_s$ the number of three-dimensional pseudovectors describing classical spins in the system, acting on vector of dimension $d$ obtained from stacking the spins into one column as $(s_1^x,s_1^y,s_1^z,s_2^x,s_2^y,s_2^z,\cdots,s_{n_s}^x,s_{n_s}^y,s_{n_s}^z)^T$, commonly referred to as the image. The matrix $\Gamma_{\text{mag}}(g)$, with $g$ being any element of the crystallographic point group, describes the combined action of the operation on the lattice sites and the magnetic moments via a reshuffling of the sites and the local rotation of magnetic moments in spin space under $g$. For example, in the honeycomb lattice with pointgroup $C_{6v}$ and two atoms in the unit cell, the rotation by $\pi$ around the $z$-axis, denoted by $C_2$, rotates both pseudovectors locally by $180^{\circ}$ and interchanges the atomic sites $A$ and $B$, resulting in the matrix:
\begin{align}
\Gamma_{\text{mag}}(C_2)=
\left(\begin{array}{cccc|ccc}
	&&A&&&B&\\
	&0&0&0&-1&0&0\\
	A&0&0&0&0&-1&0\\
	&0&0&0&0&0&1\\
	\hline
	&-1&0&0&0&0&0\\
	B&0&-1&0&0&0&0\\
	&0&0&1&0&0&0\\
\end{array}
\right).
\end{align}
The blockdiagonal form that all matrices in this representation have is obvious here.

Turning our attention now to a generic tensor we wish to expand in spin interaction terms, we require a representation of the group in higher dimensional spin space, to describe the action on terms comprising the product of any number of spins. Specifically, we calculate the invariant subspaces and corresponding eigenvalues of a projector $P$, which is the sum of all symmetry operations in the group calculated in the representation:
\begin{align}
\rho(P) = \frac{1}{|G|}\sum_{g\in G} \chi_{\rho}(g)^{\dagger}\cdot {\rho}(g),
\end{align}
where by $\rho$ we denote the representation corresponding to the order of spin $o$.
We call the eigenvalues equal to one \emph{symmetric invariants}, since they encode the spin product terms we construct in a certain order of spin in scalar values, which are invariant under operations of the crystallographic point group.

For a given tensor expressed as a product of $o$ spins, the magnetic representation induced by this tensor on the point group, $\Gamma_{\text{o}}=(\Gamma_{\text{mag}})^{\otimes o}$, will have the dimension $d^o$. Accordingly, the exponential growth of the representation's dimensionality renders the calculation of the necessary matrix operations impractical for larger system or high orders of spin.

Therefore, we make use of the fact that the product of scalars is symmetric under permutation of the scalars, and carry out the calculations in the \emph{symmetric} space, which holds only the symmetric components of the product space. This reduces the dimensionality of the resulting representation from $d^o$ to $\frac{(n+d-1)!}{n!(d-1)!}$. For example, this space does not hold all 36 components of the representation $\Gamma_{\text{mag}}\otimes \Gamma_{\text{mag}}$ for the two spin system, but only the 21 components not related by changing the order of vector components in the product. Accordingly, instead of calculating both $s_1^x\cdot s_2^y$ and $s_2^y\cdot s_1^x$, we only calculate one term, and project back from this symmetric space after the calculation. 

We thus compute the projector in the symmetric space as
\begin{align}
\rho_s(P) = \frac{1}{|G|}\sum_{g\in G} \chi_{\rho_s}(g)^{\dagger}\cdot {\rho_s}(g),
\end{align}
where by $\rho_s$ we denote the representation in the symmetric space corresponding to product space of order $o$. The symmetric invariants are now given by the eigenvalues of the invariant subspaces of this projector equal to one, hence we diagonalize the projector to find a total of $m$ such spaces, where $m$ is the multiplicity. Consequently, we find a number of $m$ such invariants for each order of spin $o$, which we denote by the symbol $I_{i,j}$, where $i\in 0,\cdots,o$ and $j\in0,\cdots,m$. 

\begin{figure}[t!]
\includegraphics[width=0.9\hsize]{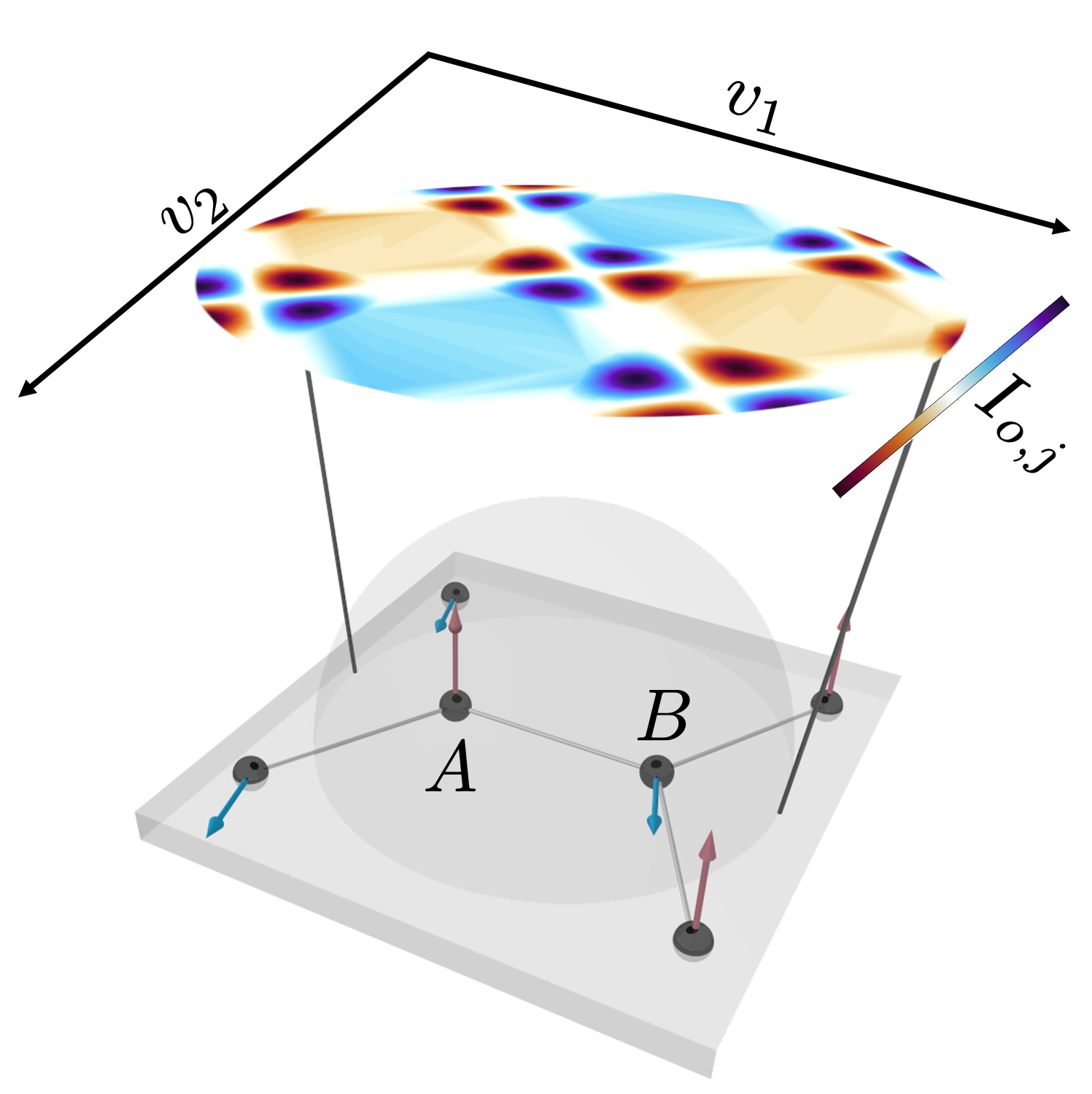}
\caption{{\bf Illustration of a two-spin invariant.} The many possible interactions between two spins on a lattice contributing to a tensor can be decoded into the symmetric invariants that the representation of the tensor induces on the point group of the lattice. It is instructive to visualize the invariants in the space of two suitable variables $v_1$ and $v_2$, for example the sum and difference of the azimuthal angles of the spins on two inequivalent sites of a honeycomb lattice $\theta_{+}=\frac{\theta_A+\theta_B}{2}$ and $\theta_{+}=\frac{\theta_A-\theta_B}{2}$. These two variables can be a good choice for a two atom system, since they correlate with the two order parameters of the magnetic configuration, which are the ferromagnetic and staggered antiferromagnetic moment $\vec{s}_{\text{FM/AFM}}=\frac{\vec{s}_A\pm\vec{s}_B}{2}$.}
\label{fig:invs_illustration}
\end{figure}

\subsection{Distribution of spin samples for two atom system}\label{sec:Sample_Dist}
We sample the magnetic moments uniformly on a sphere to cover the whole phase space offered by the magnetic configuration. The method used for this uniform sampling on a curved surface is called \emph{sphere-point picking} \cite{SPP_Weissstein}, which avoids bunching of samples near the poles when simply drawing the spherical coordinates $\varphi,\theta$ from uniform distributions. 
For $N$ moments, let $u_i$ and $v_i$, $i=1,...,N$ be random variables on $(0,1)$. Then 
\begin{align}
\nonumber   \varphi_i &= 2\pi u_i,\\
\theta_i &= \cos^{-1}(2v_i-1),
\end{align}
give the spherical coordinates with the azimuth angle $\varphi$ in the $xy$-plane, $0\leq\varphi<2\pi$, and the polar angle $\theta$ from the positive $z$-axis, $0\leq\theta\leq\pi$.

We can further augment the data computed with the TB model by applying the operations of the point group, see \cref{sec:Model_and_Syms}, on the spins and the target. By taking a look at the character table of the group, the behaviour of the target, belonging to a one-dimensional representation, can simply be deduced from the character: for the case of $A_2$, the target will pick up a sign corresponding to the character of the operation in this representation. We therefore calculate the spin configurations resulting from applying the group operations and apply the sign to the target accordingly. Additionally utilizing the known behavior of the tensor component under time reversal symmetry, we can effectively scale the size of the dataset by a factor of thirteen without the need for further TB calculations, with eleven nontrivial group operations in $C_{6v}$ and the time reversal symmetry.
\begin{figure}[t!]
\includegraphics[width=\hsize]{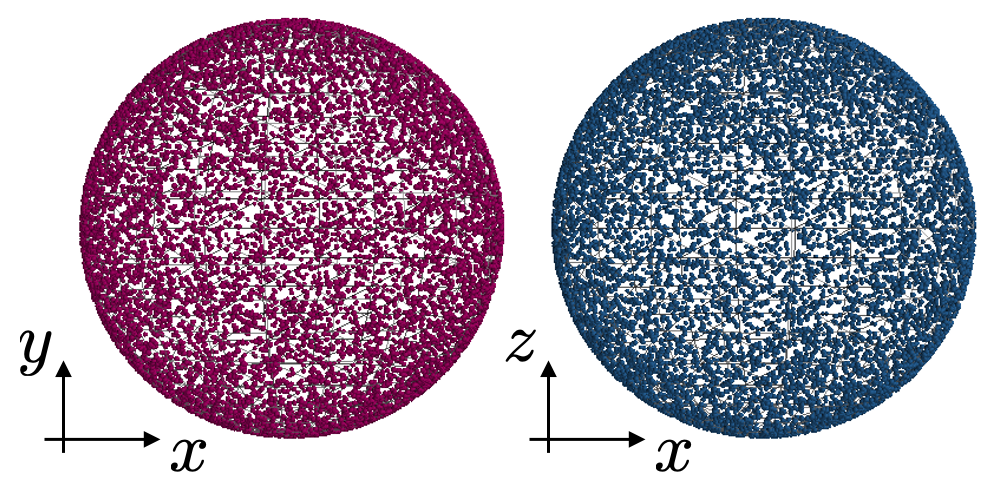}
\caption{{\bf Distribution of the directions for the two magnetic moments.} The directions of the magnetic moments are illustrated by their intersections with the unit sphere, with a view on the $xy$-plane shown in red, and a view on the $xz$-plane shown in blue. In contrast to a naive sampling of the direction, the directions cover the surface area of the unit sphere uniformly, without strong agglomeration at the poles.}
\label{fig:angledist}
\end{figure}

\subsection{Chirality and lattice site exchange}

In our following analysis, the notion of chirality $-$ specifically the vector chirality $\vec{\xi}$ defined as a vector product between spins on two inequivalent sites of the honeycomb lattice, $\vec{\xi}=\vec{s}_A\times\vec{s}_B$, see Fig.~\ref{fig:invs_illustration} $-$  and corresponding symmetric and antisymmetric in $\vec{\xi}$ contributions will play a central role.

In order to disentangle chiral and nonchiral contributions to the AHE, we (anti)symmetrize the data obtained from tight-binding calculations with respect to the rotational sense, or vector chirality $\vec{\xi}$, of the canted magnetic configuration. This symmetrization is achieved by exchange of the atomic sites $A$ and $B$, as can be easily seen from the definition of the vector chirality $\vec{\xi}$, which is antisymmetric under this operation:
\begin{align}
\vec{\xi} = \vec{s}_A \times \vec{s}_B = -(\vec{s}_B \times \vec{s}_A).
\end{align}
We therefore perform the following decomposition of the anomalous Hall conductivity tensor $\sigma_{\alpha\beta}$ $-$ or {\it any} given (scalar or tensor) quantity  of interest for that matter $-$ into even (non-chiral, $\sigma^{nc}$) and odd (chiral, $\sigma^{c}$) parts with respect to the vector chirality:
\begin{align}
\sigma_{\alpha\beta}^{c(nc)} = \frac{\sigma_{\alpha\beta}(\vec{\xi})\mp\sigma_{\alpha\beta}(-\vec{\xi})}{2}.
\end{align}
While in this work, we focus predominantly on the non-chiral part of the AHC, by seperating chiral and non-chiral parts, we aim at distinguishing chiral and non-chiral terms in the symmetrically invariant expansion as well, since this construction allows us to fit models which use $\vec{\xi}$-even or $\vec{\xi}$-odd features respectively.
Consequently, we can write the equation for the fitted model as follows:
\begin{align}
\sigma^{c(nc)} = \vec{c}_{c(nc)}\cdot\vec{I}_{c(nc)},
\end{align}
where we construct the $\vec{\xi}$-even or $\vec{\xi}$-odd invariants in the exact same way as above:
\begin{align}
\vec{I}_{c(nc)} = \frac{\vec{I}(\vec{\xi})\mp\vec{I}(-\vec{\xi})}{2}.
\end{align}
Accordingly, we make use of the symmetric and antisymmetric angles $\theta_+$ and $\theta_-$, defined as $\theta_{+}=\frac{\theta_A+\theta_B}{2}$ and $\theta_{-}=\frac{\theta_A-\theta_B}{2}$, to visualize the symmetric invariants and anomalous Hall effect and their dependence on the magnetic configuration. These two variables can be an especially good choice of latent variables for the two atom system, since they correlate with the two order parameters of the magnetic configuration, which are the ferromagnetic and staggered antiferromagnetic moment $\vec{s}_{\text{FM/AFM}}=\frac{\vec{s}_A\pm\vec{s}_B}{2}$.

\subsection{Calculation of the anomalous Hall conductivity }\label{sec:Kubo}
Given specific electronic structure, we calculate the transverse anomalous Hall  conductivity at zero temperature using the Kubo formalism, which allows us to take into account the effect of disorder in the system on the conductivity tensor. In order to do so,  we replace the retarded and advanced Green functions $G_0$ of the perfect crystal by the full Green function $G = \frac{1}{G_0^{-1}-\Sigma}$, where $\Sigma(E,\vec{k})$ is the self-energy representing the effect of disorder. Here, we use a constant broadening model such that $\Sigma(E,\vec{k})=- i\Gamma$.
With the constant broadening $\Gamma$ we obtain a Green function diagonal in the eigenspace of the Hamiltonian:
\begin{align}
G^{R/A}(E,\vec{k})_{mn}=\frac{\delta_{mn}}{E-\epsilon_{n\vec{k}}\pm i\Gamma},
\end{align}
where $\epsilon_{n\vec{k}}$ are the single-electron eigenenergies. The antisymmetric part of the conductivity tensor, which can be expressed in terms of $G^{R/A}$~\cite{CzajaAHE}, splits into two contributions:
\begin{align}
\sigma_{[\alpha\beta]}^{I}=&-\frac{1}{2\pi}\int\frac{d^3k}{(2\pi)^3}\sum_{\substack{mn\\m\neq n}}\Im\lbrace v_{mn}^{\alpha}(\vec{k})v_{nm}^{\beta}(\vec{k})\rbrace\\
\nonumber    &\times \frac{(\epsilon_{m\vec{k}}-\epsilon_{n\vec{k}})\Gamma}{((E_f -\epsilon_{m\vec{k}})^2+\Gamma^2)((E_f -\epsilon_{n\vec{k}})^2+\Gamma^2)}
\end{align}
and
\begin{align}
\sigma_{[\alpha\beta]}^{II}=&-\frac{1}{\pi}\int\frac{d^3k}{(2\pi)^3}\sum_{\substack{mn\\m\neq n}}\Im\lbrace v_{mn}^{\alpha}(\vec{k})v_{nm}^{\beta}(\vec{k})\rbrace\\
\nonumber        &\times         \frac{\Gamma}{(\epsilon_{m\vec{k}}-\epsilon_{n\vec{k}})((E_f -\epsilon_{m\vec{k}})^2+\Gamma^2)}\\
\nonumber        &-\frac{1}{(\epsilon_{m\vec{k}}-\epsilon_{n\vec{k}})^2}
\Im\left\lbrace\ln{\left(\frac{E_f-\epsilon_{m\vec{k}}+i\Gamma}{E_f-\epsilon_{m\vec{k}}+i\Gamma}\right)}\right\rbrace,
\end{align}
where $\alpha$ and $\beta$ are the Cartesian indices and $E_f$ is the Fermi energy. 
We refer to $\sigma_{\alpha\beta}^{I}$ as the Fermi-surface term in the anomalous Hall conductivity (AHC), since it only picks up contributions from the Fermi surface. The term $\sigma_{\alpha\beta}^{II}$ collects terms from all occupied states up to the Fermi level and is therefore referred to as the Fermi-sea term in the AHC.
In evaluating the Kubo expressions for the conductivity of the two-atom system we have used roughly $65\cdot 10^{4}$ $k$-points to perform the Brillouin zone integrals and a value of 25\,meV for the broadening parameter $\Gamma$.

\subsection{Algorithm layout}\label{sec:Pipeline}
In this work, we aim at fitting a linear model of the symmetric invariants $I_{o,j}$ to the target $y$,
\begin{align}
y = \sum_{o}^{o_{max}}\sum_{j=0}^{n_o} c_{o,i}\cdot I_{j}^o,
\end{align}
where $I_{o,j}$ indicates the $j$-th invariant out of a total of $n_o$ invariants in the order $o$, up to a maximal order $o_{max}$, and $c_{o,j}$ is the corresponding fitted model coefficient. Stacking all invariants and coefficients into vectors in increasing order, $\vec{x} = [x^{0}_{0},...,x^{0}_{n_0},x^{1}_{0},...,x^{1}_{n_1},...,x^{o_{max}}_{0},...,x^{o_{max}}_{n_{o_{max}}}]$, we can write:
\begin{align}
y = \vec{c}\cdot\vec{I}.
\end{align}
The symmetric invariants are possibly correlated, as are the features in many model selection tasks. Further, most of the common algorithms are optimized towards working with normal distributions with unit variance and zero mean as inputs. Therefore, we need to perform a number of tasks in order to obtain a robust feature space as a solid basis for selecting the model.

\begin{figure}[t!]
\includegraphics[width=.99\hsize]{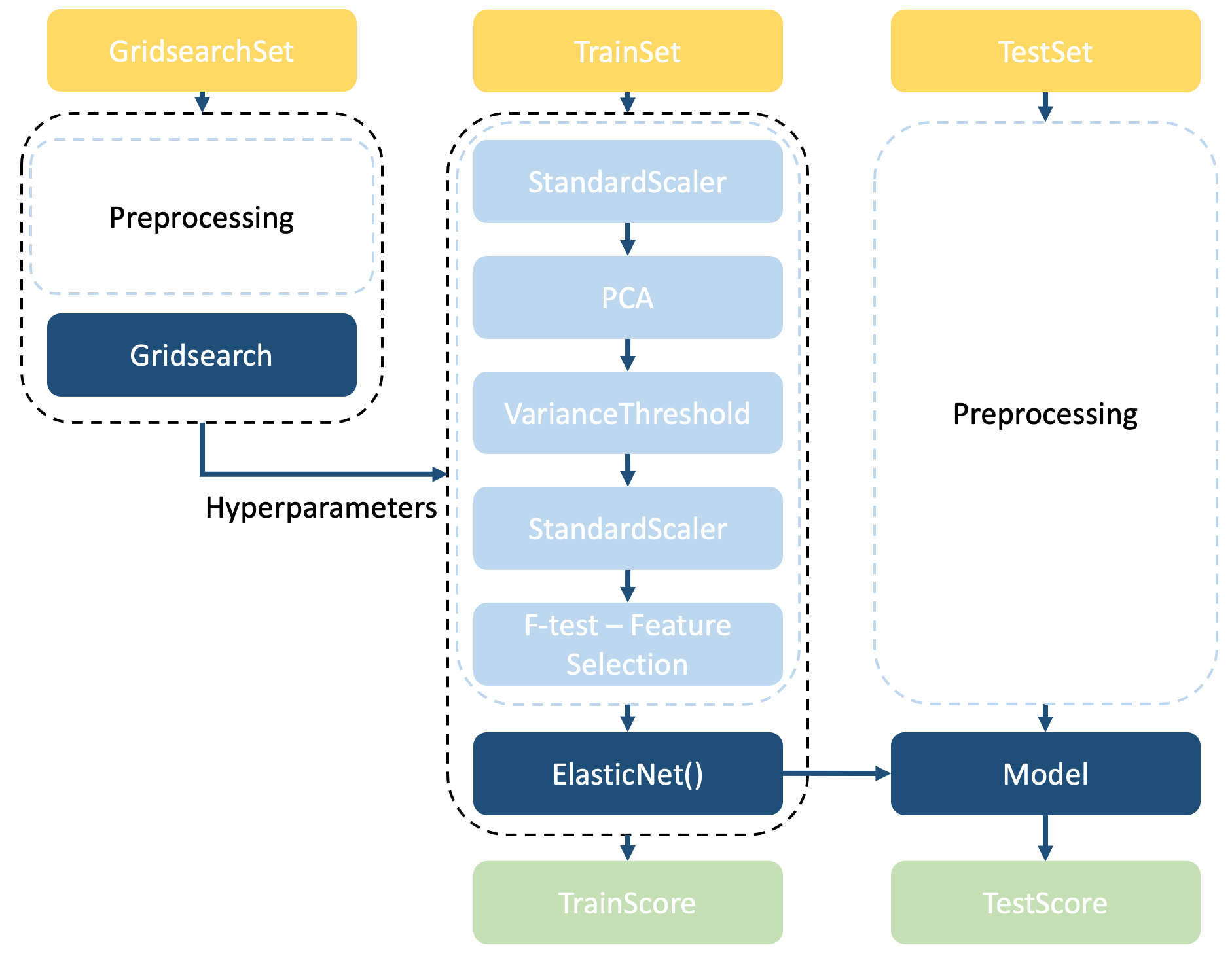}
\caption{{\bf Flowchart representation of the method.} The dataset is split into training and test sets to avoid training a model which just repeats the labels (overfitting). The training pipeline (black dashed box) consists of a number of preprocessing steps (yellow dashed box) and the model selection step. The preprocessing involves scaling the input features to unit variance and zero mean (standardization), feature decorrelation and selection by principal component analysis (PCA), see \cref{sec:PCASection}, as well as a variance threshhold and a second standardization after PCA. The hyperparameters of the pipeline are optimized via a GridSearch approach, see left column. The model selected by the pipeline is tested on a hold-out data set.}
\label{Flowchart}
\end{figure}

First and foremost, the dataset is split into training and test set to avoid training a model which just repeats the labels (overfitting). The training pipeline, see black dashed box in \cref{Flowchart}, consists of a number of preprocessing steps (light blue dashed box) and the model selection step. The preprocessing involves scaling features to unit variance and zero mean (standardization), see \cref{sec:StandardizationSection},  and decorrelation and feature selection by principal component analysis (PCA), see \cref{sec:PCASection}, as well as a variance threshold removing all features with variance below a given threshold and standardization after PCA.
We then perform the model selection step by training regularized models on the features selected by the PCA method, see \cref{ModelSelSection}. We take care to avoid data leakage along the pipeline: if we perform e.g. the PCA on the whole dataset before splitting into train and test set, the prediction of the model will be overly optimistic. Therefore, the preprocessing transformations are trained on the training set only.

\subsubsection{Standardization}\label{sec:StandardizationSection}

The model selected by cross validation maps the input features $\vec{x}_m$ onto the single input label $y_m$ via the model coefficients $\vec{c}_m$, where the $M$ true features are obtained from $N$ observations:
\begin{align}\label{modelregression}
y_m = \vec{c}_m^T\cdot\vec{x}_m.
\end{align}
Input features and labels have been obtained from the true features and labels by standardization, subtracting the mean $\mu$ and dividing by the standard deviation, or scale, $s$, before training the model:
\begin{align}\label{relations}
&y_t^j = s_y\cdot y_m^j + \mu_y,\\
&\vec{x}_t^j = \myMat{D}_{[\vec{s_x}]}\cdot\vec{x}_m^j + \vec{\mu}_x,\\
&\vec{\mu}_x = \frac{1}{N}\sum_j \vec{x}_t^j=\frac{1}{N}\sum_j (\myMat{D}_{[\vec{s_x}]}\cdot\vec{x}_m^j + \vec{\mu}_x),\\
&\mu_y=\frac{1}{N}\sum_j y_t^j=\frac{1}{N}\sum_j (\vec{c}_t^T\vec{x}_t^j)=\vec{c}_t^T\vec{\mu}_x,
\end{align}
where $\myMat{D}_{[\vec{s_x}]}$ is the matrix containing the standard deviation of each true feature on the diagonal, $\myMat{D}_{[\vec{s_x}]}^{ii}=s_x^i$, with all other entries equal to zero. The vector $\vec{\mu}_x$, contains the mean values of each true feature, and the summation index $j$ runs over the samples.

\subsubsection{Principal component analysis}\label{sec:PCASection}

Further, the standardized features are decorrelated by using \emph{principal component analysis} (PCA) \cite{Szlam2014_PCA}. This method finds the singular value decomposition of the feature matrix, which is in essence a generalization of finding the diagonal of the feature matrix. The method of singular value decomposition, on which PCA is based, relies on a following geometric consideration~\cite{Trefethen1997_NumericalLinAlg}. Considering a unit sphere in $n$ dimensions, the image of this unit sphere under any $m\times n$ matrix is a hyperellipse. A hyperellipse is the $m$ dimensional equivalent of the twodimensional ellipse, obtained by stretching the twodimensional unit circle by some factors $\sigma_1,\dots,\sigma_m$ along some orthogonal directions $u_1,\dots,u_2\in\mathbb{R}$, which are, for convenience, normalized to unit length, $||u_i||_2=1$. We refer to the vectors ${\sigma_i u_i}$ as the principal semiaxes of the hyperellipse, with length $\sigma_i$, and for a matrix $A$ of rank $r$, exactly $r$ of these lengths turn out to be nonzero. In particular, if $m\geq n$, at most $n$ of the will be nonzero. 

With this geometric picture in mind, let us define the singular value decomposition (SVD) of $A\in\mathbb{C}^{m\times n}$: let $m$ and $n$ be arbitrary natural numbers $\geq 0$, then the SVD of $A$ is defined as:
\begin{align}
A=U\Sigma V^{\ast},
\end{align}
where 
\begin{align}
\nonumber	U\in\mathbb{C}^{m\times m}\text{ is unitary,}\\
\nonumber	V\in\mathbb{C}^{n\times n}\text{ is unitary,}\\
\Sigma\in\mathbb{R}^{m\times n}\text{ is diagonal.}
\end{align}
Additionally, we assume the diagonal entries $\sigma_j$ of $\Sigma$ to be non-negative and ordered in nonincreasing order, $\sigma_1,\geq\dots,\sigma_p\geq0$, where $p=min(m,n)$. Looking at this definition, above statement is made quite clear: the unit sphere $S\in\mathbb{R}^n$ is transformed to the hyperellipse $SA\in\mathbb{R}^m$, since the unitary map $V^{\ast}$ preserves the sphere, the diagonal matrix $\Sigma$ stretches it and the unitary matrix $U$ rotates and reflects the ellipse. For a proof, that every matrix $A$ has such an SVD, see also Ref.~\cite{Trefethen1997_NumericalLinAlg}.

The usefulness of the SVD for machine learning problems can now be illustrated by investigating the diagonal matrix $\Sigma$. Considering the SVD of a feature matrix of shape $m\times n$, constructed from stacking the vectors of $n$ features for all the $m$ samples, the matrix $\Sigma$ is potentially very large, especially for problems with many samples, many features, or both. 
However, two characteristics of this matrix make the SVD worth calculating. On one hand, the fact that $\Sigma$ is diagonal implies that the corresponding singular vectors are orthogonal, or in statistical terms, uncorrelated. On the other hand, the number of non-trivial singular values is not given by the dimensions $m$ or $n$, but by the rank $r$ of the matrix, which can be significantly smaller than $m$ and $n$ for features that are correlated. Accordingly, for matrices with rank $r<n,m$, the resulting feature space is potentially much smaller than the original feature space.

In the context of machine learning, methods involving singular value decomposition are referred to as principal component analysis (PCA) methods, where the principal components are equivalent to the coordinates of the principal axes, and the measure of explained variance corresponds to the singular value, or length, of the principal axes. This correspondence lends itself to the analysis of variance in the feature space, as the singular values, as the norm of the singular axis they correspond to, determine how much overall feature variance a change along a certain axis induces.

\subsubsection{Variance Threshold}\label{VarTreshSection}

Features with low variance will very likely not have a large influence on the model target. To compare feature variances in a meaningful way, we first scale all the features to the range $[0,1]$ individually by using a MinMaxScaler. We then discard features with a variance below a certain threshold. This allows us to reduce the dimension of the feature space with very small effort. The threshold for the data shown in this work is $v_{t}=10^{-5}$.

\subsubsection{Feature selection with statistical methods}

Since the modelling technique introduced in this work relies on an expansion agnostic of any physical intuition regarding the presence of certain interactions like exchange or spin-orbit coupling in the model Hamiltonian \cref{eq:model}, condensation of the input feature space to the relevant components is a key ingredient for meaningful modelling of the electric transport.

To this end, several techniques can be utilized in conjunction with the regularized regression algorithm, which prefers a sparse description of the target by penalizing models with many nonzero coefficients. While the PCA-transformation can be used to extract portions of the latent space explaining overall variance of the feature space, we suggest to use a statistical measure which incorporates also the correlation with the target variable.

Specifically, we focus here on a technique estimating the linear correlation between single input features and the target variable, the \emph{f-regression test}, since the working hypothesis is that the electric transport can be described by a linear model of the input features.

This quantity measures the cross-correlation between two variables $x$ and $y$ over $N$ observations, given by the Pearson correlation coefficient:
\begin{align}
\nonumber    R &= \frac{\text{cov}(x,y)}{\sigma_{x}\sigma_{y}}\\
&=\frac{\sum_{i}^N (x_i-\bar{x})(y_i-\bar{y})}
{\sqrt{\sum_{i}^N(x_i-\bar{x})^2}\sqrt{\sum_{i}^N(y_i-\bar{y})^2}}.
\end{align}
Internally, the Pearson correlation coefficient is converted to an F-score (a number between 0 and 1). Computing the f-regression score for each component and ranking the components according to this metric, we can select a top scoring percentile to fit the model and disregard features scoring below a threshold. This approach nicely complements the PCA transformation step, which ranks the constructed components by explained variance ratio in the feature space. The f-regression score presents a systematic way to assess the importance of input features for the model based on their correlation with the target, which the PCA transformation is not able to do.

\subsubsection{Model selection by regularized regression}\label{ModelSelSection}
We regularize this model by the \emph{elastic net} penalty \cite{Friedman2010_Lasso}. The regularized loss-function $F(w)$ to minimize is therefore the sum of loss-function $L(\vec{c}^T\cdot \vec{x},y)$ and regularization term $R(\vec{c})$:
\begin{align}\label{eq:elnetloss}
\nonumber	F(c) =& L(\vec{c}^T\cdot \vec{x},y)+\alpha\cdot R(\vec{c})\\
\nonumber	=& \frac{1}{ 2 N} \cdot ||y - \vec{I}\cdot\vec{c}||^2_2\\
\nonumber&+ \alpha \cdot l_1 ||\vec{c}||_1\\
&+\frac{1}{2} \alpha \cdot (1 - l_1) ||\vec{c}||^2_2,
\end{align}
where $N$ is the number of samples, and $\alpha$ is the parameter controlling the strength of regularization, with $\alpha=0$ corresponding to an ordinary least squares (OLS) fit. The OLS fit can be adapted to an \emph{squared-epsilon-insensitive} loss, by allowing the model to ignore errors below a certain threshold of $\epsilon$, and applying the standard squared loss above the threshold. This adapted error function allows for slightly improved description of outliers in some cases. We use a value of $\epsilon=10**{-2}$ here.

The parameter $l_1$ is the ratio interpolating between two kinds of regularization related to a norm on the weights or coefficients of the model, LASSO and RIDGE, by $L_1$-norm ($l_1=1$, LASSO) or by $L_2$-norm ($l_1=0$, RIDGE). While the RIDGE regression penalizes the absolute size of coefficients, LASSO penalizes the number of nonzero coefficients. In general, RIDGE regression gravitates towards models where the size of all coefficients is optimal. In contrast, LASSO prefers sparse models, where the number of nonzero coefficients is optimal.

Regularization as a mix of the two methods helps us to find a sparse model, with the optimal number of nonzero coefficients, while keeping the size of coefficients in check, and is also commonly referred to as \emph{compressive sensing}. The models are ranked by a scoring method of choice, in this case the coefficient of determination, $R^2$, calculated from the residual sum and total sum of squares:
\begin{align}\label{eq:coeffdet}
\nonumber   u&=\sum_{i}^N(y_{\text{true}}^i-y_{\text{predict}}^i)^2,\\
\nonumber   v&=\sum_{i}^N(y_{\text{true}}^i-\bar{y}_{\text{true}})^2.\\
R^2 &= (1-\frac{u}{v}).
\end{align}
Here, for a sample numbered by the index $i$, we denote the true value of the target by $y_{\text{true}}^i$ and the model prediction for the sample by $y_{\text{predict}}^i$. The mean of all true target values is calculated as $\bar{y}_{\text{true}}=1/N\sum_{j}^N y_{\text{true}}^j$, where $N$ is the total number of samples.

The best possible score is $1.0$, and the score can be negative, since the model can be arbitrarily bad. The score on either train or test set is not our only measure of the models fidelity: we can additionally probe the functional dependence along specific paths in the feature space. This allows us to quantify not only statistical quantities like the MSE, but also to probe how faithfully the model reproduces the functional dependence of the target along certain paths or in certain regions, for example varying only the polar angle $\theta$ on an arc starting from the north pole, $\theta\in\{0,\pi\}$, keeping the rest the same. Especially when the model extrapolates from one region, overfitting in this regard must be closely monitored.

\subsubsection{Stochastic Gradient Descent regression}
The regressor is the tool that minimizes the loss function defined by the regularization, elastic net defined in \cref{eq:elnetloss} in this case, on the given training data. We use the Stochastic Gradient Descent (SGD) regressor here, which is an especially effective optimization tool for large quantities of training data, usually $>10^{4}$ samples. SGD estimates the true gradient of the loss function on one training sample, as opposed to other methods like batch gradient descent, and updates the weights along the gradient. One pass over the whole training set is referred to as one iteration here. After an estimate of the gradient $\partial F(\vec{c})/\partial\vec{c}$ has been computed on one sample, the model coefficients, or weights, are updated as follows:
\begin{align}\label{eq:SGD}
\nonumber\vec{c} &\leftarrow \vec{c}-\mu\frac{\partial F(\vec{c})}{\partial \vec{c}}\\
\vec{c} &\leftarrow \vec{c}-\mu\left[\alpha\frac{\partial R(\vec{c})}{\partial \vec{c}}+\frac{\partial L(\vec{c}^T\cdot \vec{x},y)}{\partial \vec{c}}\right],
\end{align}
where by $R(\vec{c})$ we refer to the regularization term of the elastic net and by $L(\vec{c}^T\cdot \vec{x},y)$ we refer to the loss-function, as defined in \cref{eq:elnetloss}, and $\mu$ is the learning rate.

\subsubsection{Hyperparameter optimization via cross validation}
SGD is influenced by a number of hyperparameters, which we choose based on a cross-validation search. This method is used to optimize the values of hyperparameters for a particular algorithm by repetitively running the algorithm for different parameter values and thereby determining the highest scoring point in the parameter space. Note that the score here does not necessarily need to be the test score of the resulting model found by the algorithm, but can be any metric supplied by the user. Some obvious examples would be to optimize for minimal runtime, optimal model score, or even a trade off between the two. We make use of this tool to finetune the regularization parameter $\alpha_{\text{reg}}$, and to find the optimal $l_1$-ratio between Lasso and Ridge in the elastic net, as well as the optimal learning rate $\mu$. In general, different loss functions can be probed in order to better handle outliers or zero inflation, like the \emph{squared epsilon insensitive} loss function, which ignores errors below a given epsilon.

In order to achieve the highest efficiency for the cross validation method, we employ the \emph{Halving Grid Search} algorithm. A given set, or grid, of parameters, serves as candidates. All candidates are evaluated in the first pass with a limited amount of resources, and a top fraction of the candidates with respect to a scoring method of the users choice is kept while the rest is discarded. The surviving candidates are then evaluated again with an increased amount of resources and the process is repeated.
While the number of candidates carried over to the following iterations is divided by factor $s$, the amount of resources is multiplied by the same factor $s$.

A hyperparameter search using this method is especially efficient, since not all candidates are evaluated on equally large amounts of resources, in this case training samples. In order to avoid overfitting, the algorithm employs cross validation as well by using the $k$-fold strategy, where the available resources are split into $k$ consecutive folds, each of the used as a validation set once while all other folds form the training set.

\subsubsection{Backtransformation of model coefficients}
The model coefficients are not the true physical coefficients connecting true physical features and labels.
However, to interpret the trained model in physical terms, we require the coefficients in the physical feature space:
\begin{align}\label{trueregression}
	y_t = \vec{c}_t^T\cdot\vec{x}_t.
\end{align}
A mapping between model coefficients and true coefficients is found by applying the inverse transformation along the pipeline to the coefficients. This means inserting \cref{modelregression} and \cref{relations} from \cref{sec:StandardizationSection} into \cref{trueregression}, and multiplying by $ \vec{x}_m^T\cdot\myMat{D}_{[\vec{s_x}]}^{-1}$, to arrive at the coefficients for the PCA-transformed features $\vec{x}_p$: 
\begin{align}
	\nonumber   &\vec{c}_p^T\cdot\vec{x}_p=s_y\cdot(\vec{c}_m^T\cdot\vec{x}_m)+\mu_y\\
	\nonumber   \Rightarrow&\vec{c}_p^T\cdot\myMat{D}_{[\vec{s_x}]}\cdot\vec{x}_m=s_y\cdot(\vec{c}_m^T\cdot\vec{x}_m)+\mu_y-\vec{c}_t^T\vec{\mu}_x\\
	\Rightarrow&\vec{c}_p=s_y\cdot(\vec{c}_m^T\cdot\myMat{D}_{[\vec{s_x}]}^{-1}).
\end{align}
Then, applying the inverse transformation of the PCA, we arrive at the true coefficients of the original symmetric invariants:
\begin{align}
	\vec{x}_t=P^{-1}(\vec{x}_p),
\end{align}
We can therefore extract the physical parameters from the trained model by multiplying with the standard deviation of the target and dividing by the standard deviation of the feature to reverse a standardization step or by applying the inverse transform of the exactly invertible transformations along the pipeline, such as PCA. 

\FloatBarrier
\section{Results}\label{sec:PipelineResults}
\subsection{Input data}

By making use of the method presented in \cref{sec:Sample_Dist}, we generate a total number of $10^4$ samples, each describing the two magnetic moments on sites $A$ and $B$ of the bipartite honeycomb lattice. The moments point in randomly chosen directions, such that the overall distribution of directions is uniform over the unit-sphere, see Fig.~\ref{fig:angledist}.
The anomalous Hall conductivity, which represents the target quantity, is numerically calculated by using the Kubo approach detailed in \cref{sec:Kubo}. The calculation provides the target value for different values of the Fermi energy, which allows the analysis of the fitted models with respect to changes when probing the transport at different points in the electronic structure, characterized by different regimes of reciprocal geometry.

An explicit implementation of the symmetry expansion, outlined in \cref{sec:Expansion}, provides input features to the modelling pipeline. Since the anomalous Hall conductivity tensor contains an odd number of spins, the even expansion terms for the anomalous Hall conductivity tensor vanish. 
In principle, objects of arbitrary order in spin can appear in the model. For the sake of practicability, we calculate the first six non-vanishing orders of the expansion, namely orders $\{1,3,5,7,9,11\}$ with $\{1,7,26,76,185,392\}$ terms respectively, representing a complete description of possible interaction terms up to order eleven in spin, containing $687$ different terms. We clarify specifically, that higher orders in spin do not incorporate a higher number of distinct magnetic moments. The order in spin corresponds plainly to the power of the image, which is the pseudovector of dimension $d$ introduced in \cref{sec:Expansion}, representing the magnetic state of the system.

\begin{figure}[t!]
	\includegraphics[width=0.9\hsize]{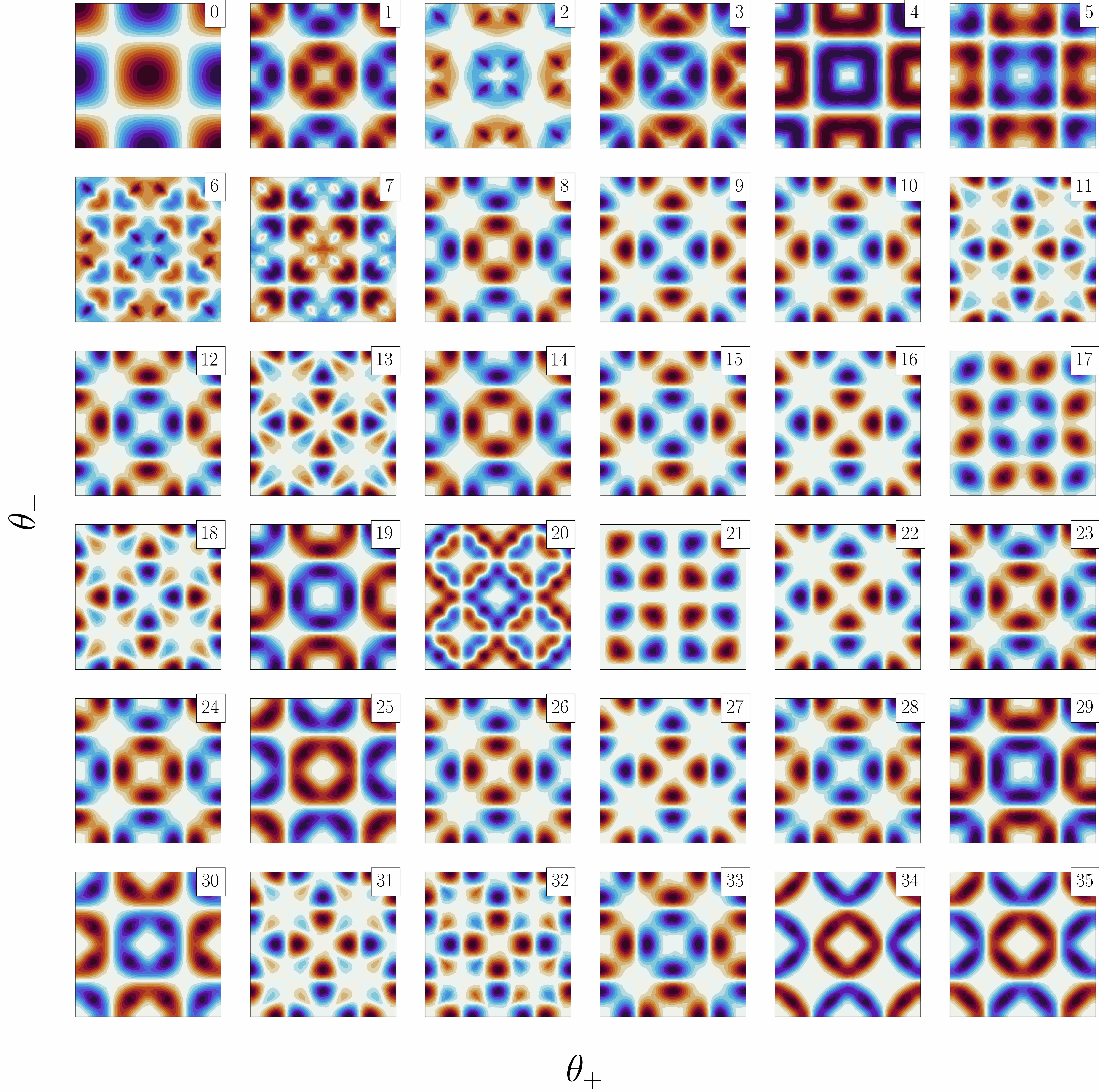}
	\caption{{\bf Invariants symmetrized with respect to exchange of atomic sites.} Invariants are shown one per panel, with order increasing from top left to bottom right, as functions of $\theta_+$ and $\theta_-$, with both variables spanning the range $[0,\pi]$. The center of the plot is located at the coordinates $(0,0)$. In this parameter space the invariants show distinct features, which allows to categorize them into a handful of classes. 
	}
	\label{fig:sym_invs}
\end{figure}
\begin{figure}[t!]
	\includegraphics[width=0.9\hsize]{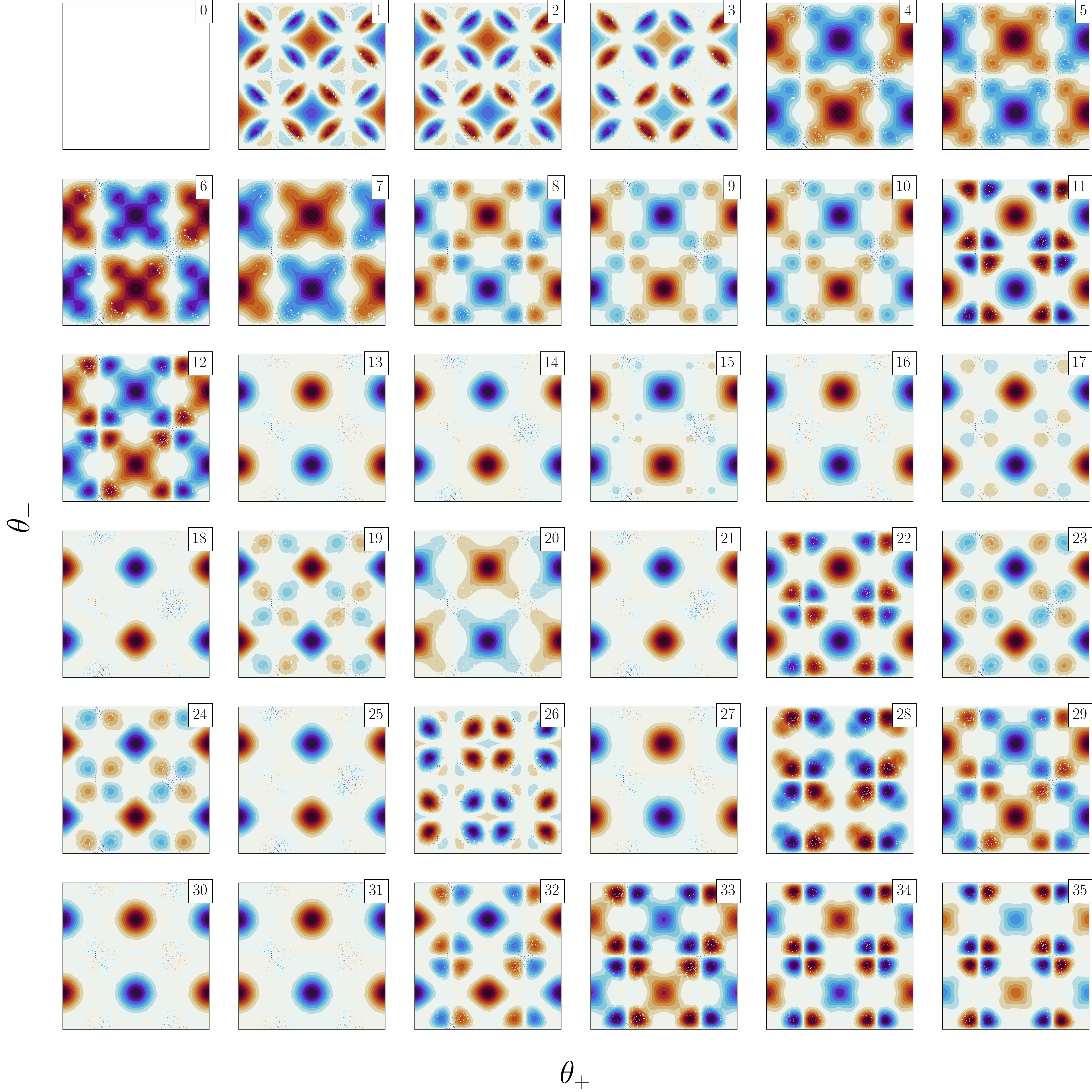}
	\caption{{\bf Invariants antisymmetrized with respect to exchange of atomic sites.} Invariants are shown one per panel, with order increasing from top left to bottom right, as functions of $\theta_+$ and $\theta_-$, with both variables spanning the range $[0,\pi]$. The center of the plot is located at the coordinates $(0,0)$. For the discussion of morphologies, see text. 
	}
	\label{fig:asym_invs}
\end{figure}

In  Figs.~\ref{fig:sym_invs} and \ref{fig:asym_invs} we present an overview of the symmetric invariants calculated for the two-atom system. They show the original features symmetrized and antisymmetrized with respect to vector chirality $\vec{\xi}$ respectively, the panels show one invariant each as a function of azimuth angles $\theta_+$ and $\theta_-$, sorted by increasing expansion order from top left to bottom right.
Inspecting the data shown in Fig.~\ref{fig:sym_invs}, the invariants reveal their structure as functions of the spherical coordinates. All $\vec{\xi}$-even invariants share the more or less sharp nodal lines in both horizontal and vertical direction, at coordinates $\theta_{+}\in n\cdot\pi+\pi/2, n\in \mathbb{N}$ or $\theta_{-}\in n\cdot\pi+\pi/2, n\in \mathbb{N}$. Between these nodal lines, blocks with alternating sign are present, which grow in complexity for higher order, displaying multiple sign changes (panel $6$), polar features (panel $16$) and fine modulations (panel $11$). Investigating the data presented in the first panel, displaying the most simple structure, the correlation with the overall magnetization, or sum of the two magnetic moments of the system, is evident. This in turn promotes the idea, that this first invariant should be assigned large weight in the description of the anomalous Hall effect, since this phenomenon is conventionally associated with the overall magnetization. Invariants beyond the $36$ shown here display quite similar morphology, with the same building blocks repeating with slight variations.

The interpretation of these patterns becomes more intuitive  when associating the values of $\theta_{\pm}$ with specific magnetic configurations. In the center of the panels, at coordinates $(0,0)$, we find the standard ferromagnetic alignment of moments along the $z$-axis in positive direction. Traversing the panel in the horizontal direction corresponds to a collective rotation of both spins around an axis, with the nodal lines indicating the moments passing through the $xy$-plane and the AHE consequently vanishing. A similar explanation holds for the vertical direction, and traversing along this axis corresponds to a rotation by equal amounts in different directions per site, which we refer to as spin canting. However, the nodal lines along the $\theta_{-}$-axis are not associated with the moments passing through the $xy$-plane, but rather with the moments falling into an antiferromagnetic alignment. When either rotation or canting angle has reached the value of $\pi$, the magnetic configuration has been reversed to the ferromagnet aligned with the $z$-axis in negative direction.

If we imagine now, that we only rotate one of the spins around an axis in the $xy$-plane, then both $\theta_{+}$ and $\theta_{-}$ will increase or decrease with the same rate. In this way, depending on the rotation direction, we reach the four nodes placed at the corners of the tiles. These correspond to the antiferromagnetic alignment along the $z$-axis, with one spin aligned in positive and negative $z$-direction each. As can be easily imagined, traversing from here in the $\theta_{+}$ direction mimics the collective rotation of the antiferromagnet around an axis in the $xy$-plane, with the nodal lines again representing the position where the moments pass through this plane.

Focussing now on the $\vec{\xi}$-odd invariants shown in Fig.~\ref{fig:asym_invs}, the most obvious feature of the data presented here is that the first invariant is exactly zero everywhere. Considering the fact that this invariant is the only one of first order in spin, it is conceivable that this invariant is related to the overall magnetization. As the magnetization, or ferromagnetic moment of the configuration, is insensitive to exchange of the lattice sites, the antisymmetrization leads to a cancellation.
As is the case of $\vec{\xi}$-symmetric invariants, there are dominant nodal lines, however at different coordinates given by $\theta_{+}\in n\cdot\pi+\pi/2, n\in \mathbb{N}$ or $\theta_{-}\in n\cdot\pi, n\in \mathbb{N}$. These nodal lines transform into quite different morphologies in higher order, namely the kind similar to the data shown in panel $11$, with quadrupoles surrounding the critical points of $(\theta_{+},\theta_{-})\in {(n\cdot\pi+\frac{\pi}{2},m\cdot\pi))}, n\in \mathbb{N},m\in\mathbb{N}$ and isotropic features at coordinates $(\theta_{+},\theta_{-})\in {(n\cdot\pi,m\cdot\pi+\frac{\pi}{2}))}, n\in \mathbb{N},m\in\mathbb{N}$. Additionally, only the isotropic characteristics are present, e.g. in panels $11,13,14,15,16$ at points shifted by $\pi/2$ from the nodal points mentioned before.

Referring to explicit magnetic arrangements, we can relate the characteristics of the invariants to specific properties of the magnetic configuration. The horizontal nodal line, visible in all of the antisymmetric invariants, marks the ferromagnetic configurations with $\theta_-=0$, connected by collective rotation of spins. The two vertical nodal lines then correspond to the arrangements with ferromagnetic moment rotated into the $xy$-plane. In contrast to the symmetrized data, the antiymmetrized components are all exactly vanishing for perfect ferromagnetic alignment of spins, but not for the antiferromagnetic arrangement, with the exception of coordinates $(\pm\pi/2,\pm\pi/2)$. These coordinates mark an antiferromagnetic arrangement in the $xy$-plane. The isotropic features, appearing at coordinates $(\theta_{+},\theta_{-})\in {(n\cdot\pi,m\cdot\pi+\frac{\pi}{2}))}, n\in \mathbb{N},m\in\mathbb{N}$, correspond to the perfect antiferromagnetic configurations along the $z$-axis, perpendicular to the lattice plane.

Finally, the dataset is augmented by applying the lattice symmetries, as detailed in \cref{sec:Model_and_Syms}. From each of the spin configurations, $12$ additional configurations can be generated, one for each nontrivial group operation and one for the time reversal operation. Multiplication with the operations character in the representation yields the corresponding values of the anomalous Hall conductivity tensor and the symmetric invariants from the original values.

\begin{figure}[t!]
	\includegraphics[width=0.9\hsize]{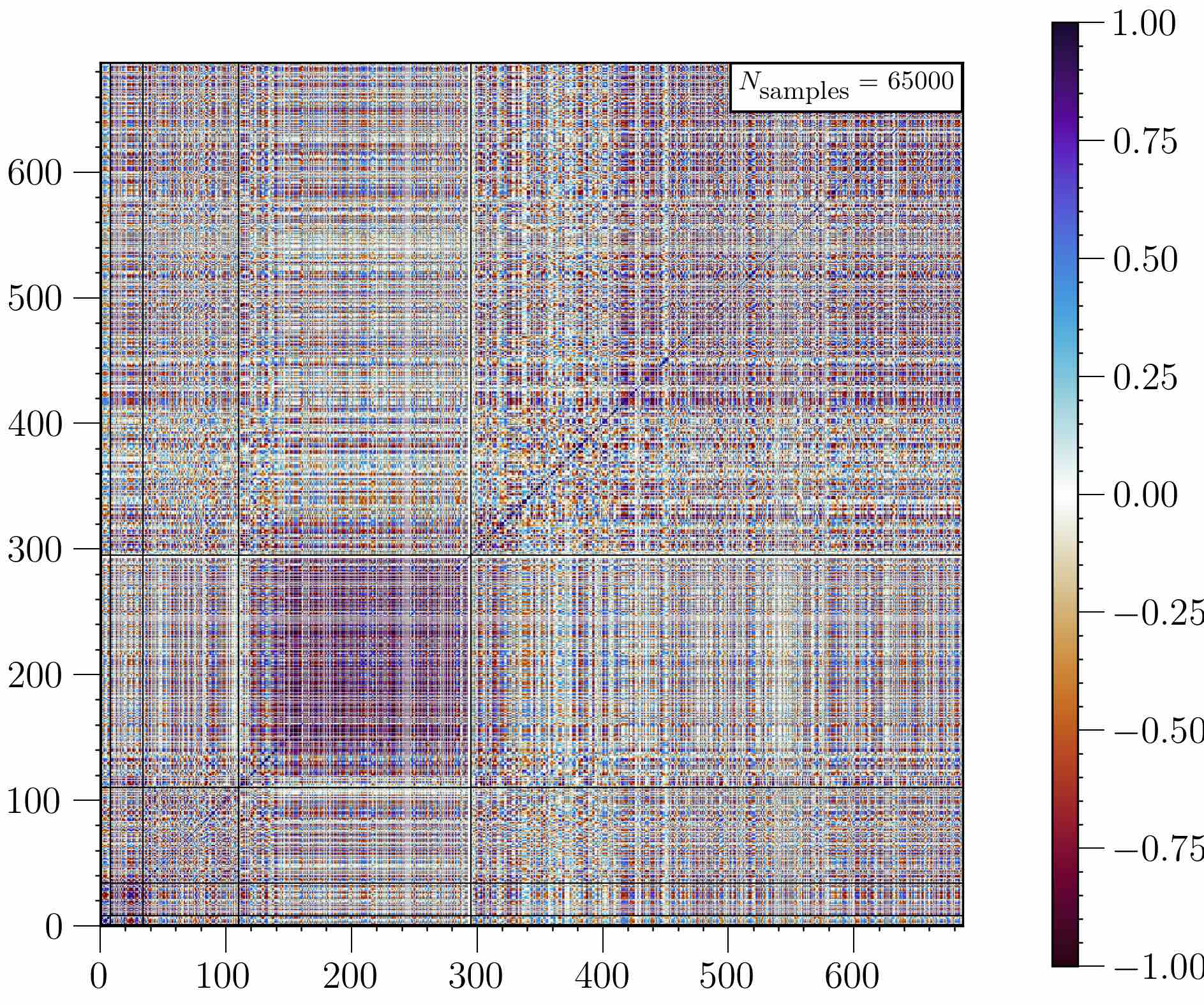}
	\caption{{\bf Cross-correlation matrix of the input features.} The correlation matrix of symmetric invariants symmetrized with respect to vector chirality $\vec{\xi}$. The cross-correlation is calculated over the total number of samples, which is $65000$. Obviously, the symmetric invariants are, as already apparent from the similar structure in the space of $(\theta_{+},\theta_{-})$, heavily correlated amongst each other. Perfectly uncorrelated features would result in a matrix with ones along the diagonal and zeros everywhere else.}
	\label{fig:corr}
\end{figure}
\begin{figure}[t!]
	\includegraphics[width=0.90\hsize]{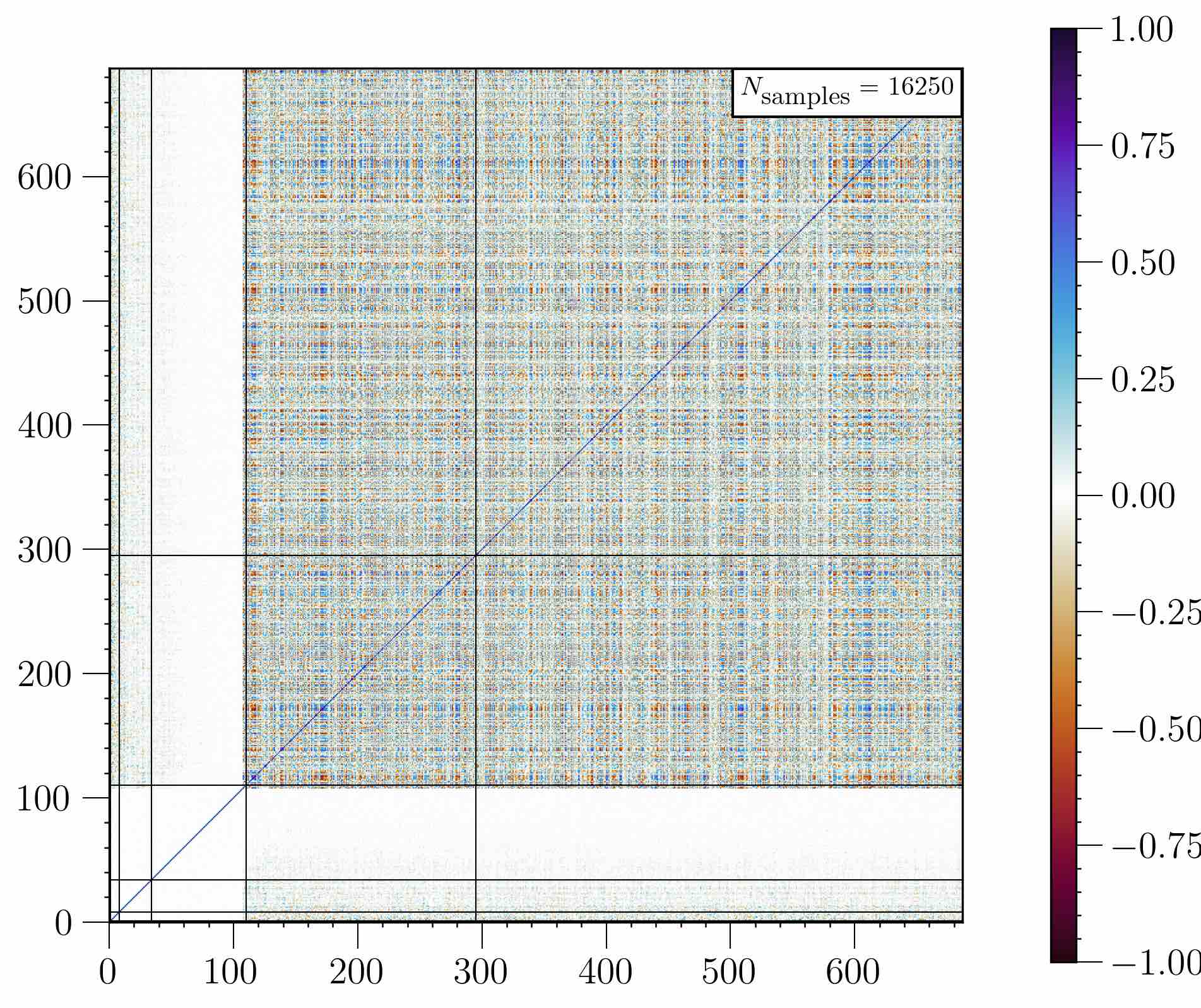}
	\caption{{\bf Result of principal component analysis (PCA).} The correlation matrix of symmetric invariants symmetrized with respect to vector chirality $\vec{\xi}$, after principal component analysis (PCA). The total number of samples used here is $65000$, the PCA transformation is trained on $44750$ samples and $16250$ samples are held out as a test set, shown in the top right corner. The correlation matrix assumes a block diagonal form after performing the transformation found by PCA. 
	}
	\label{fig:PCA_corr}
\end{figure}

In conclusion, the modelling pipeline works on a set of $13\cdot10^4$ samples $-$ obtained from augmenting the $10^{4}$ initial TB calculations by exploiting the cyrstal symmetries and time reversal symmetry $-$ where each sample is characterized by a magnetic configuration of $\vec{s}_A$ and $\vec{s}_B$, a corresponding value of the anomalous Hall conductivity, and a collection of $687$ symmetric invariants.
Before continuing with any further steps, the feature data are standardized by substracting the mean and scaling by the variance, as illustrated in \cref{sec:Pipeline}.

\subsection{PCA transformation}

After standardizing the features according to \cref{sec:Pipeline}, a PCA transformation constructs a decorrelated and greatly condensed feature space, as clarified in \cref{sec:PCASection}. This transformation is especially useful for the kind of data considered here, as the features show strong correlation, see off-diagonal terms in \cref{fig:corr}.
Turning our attention to the data shown in \cref{fig:PCA_corr}, we realize that the correlation matrix assumes a block diagonal form after performing the transformation found by PCA. The matrix displays a square block in the bottom left corner, indicating that the PCA transformation finds a number of $\sim 110$ uncorrelated components in the feature space. While the amount of off-diagonal weight in this matrix is visibly less than before, the off-diagonal terms are quite large, especially in higher orders, for feature indexes above $110$. Closer inspection of the components in this region reveals that the explained variance of the components is orders of magnitude smaller than for the first components, and therefore the cross-correlation is inflated by the inverse dependence on the variance scale. Accordingly, we discard features beyond $\sim110$ by applying a variance threshold after the PCA transformation.

\begin{figure}[t!]
	\includegraphics[width=0.9\hsize]{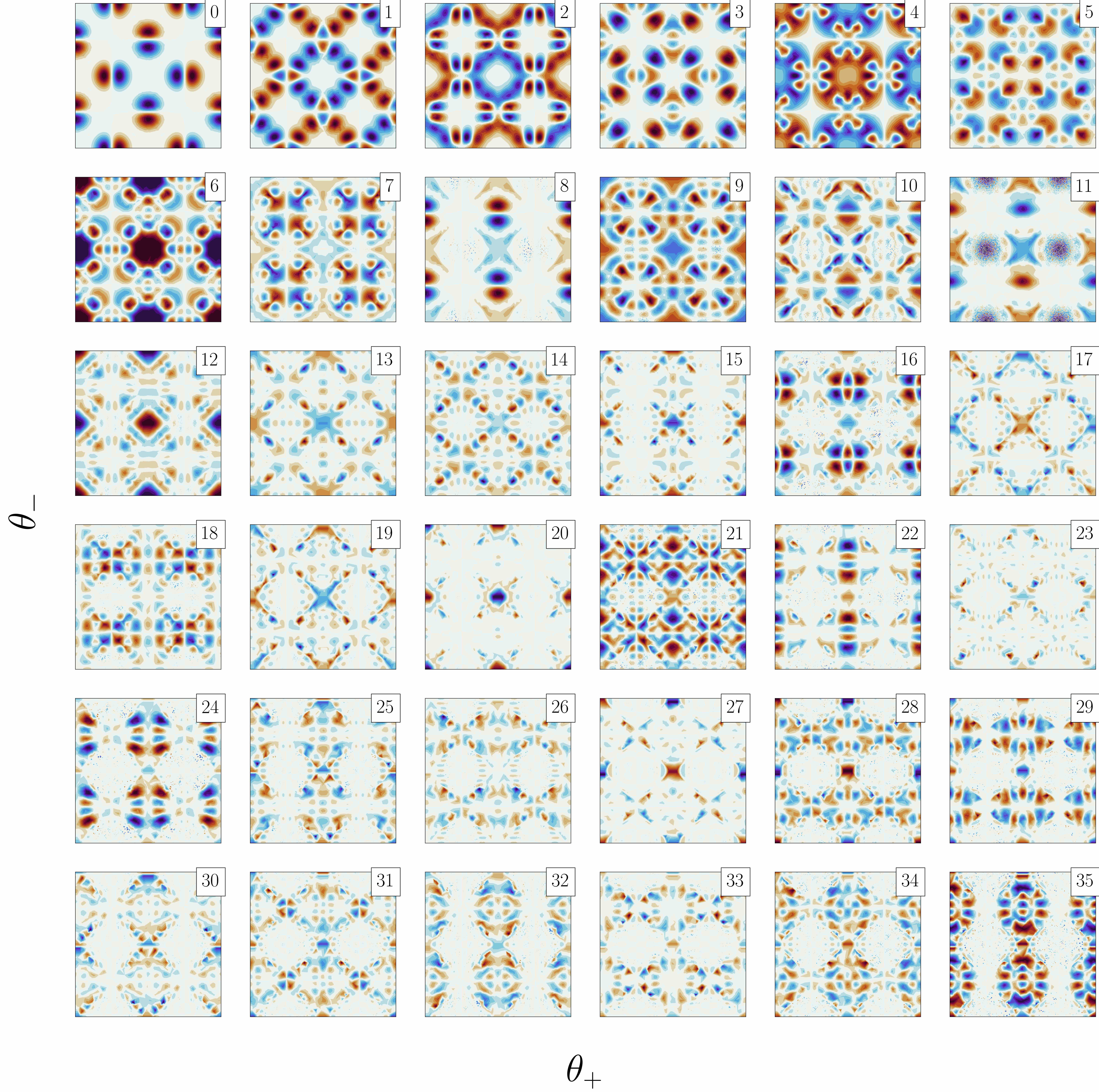}
	\caption{{\bf PCA-transformed invariants symmetrized with respect to exchange of atomic sites.} 
		The components are normalized to one in each panel, otherwise higher order features would not be visible due to their small scale. The transformation effectively merges by linear combination the multiple, quite similar symmetric invariants shown in Fig.~\ref{fig:sym_invs}, into a few relevant morphologies as the dominant directions in the resulting latent space with respect to explained variance. This allows to reduce the input feature space to a handful of relevant features.}
	\label{fig:sym_invs_pca}
\end{figure}

\begin{figure}[t!]
	\includegraphics[width=0.9\hsize]{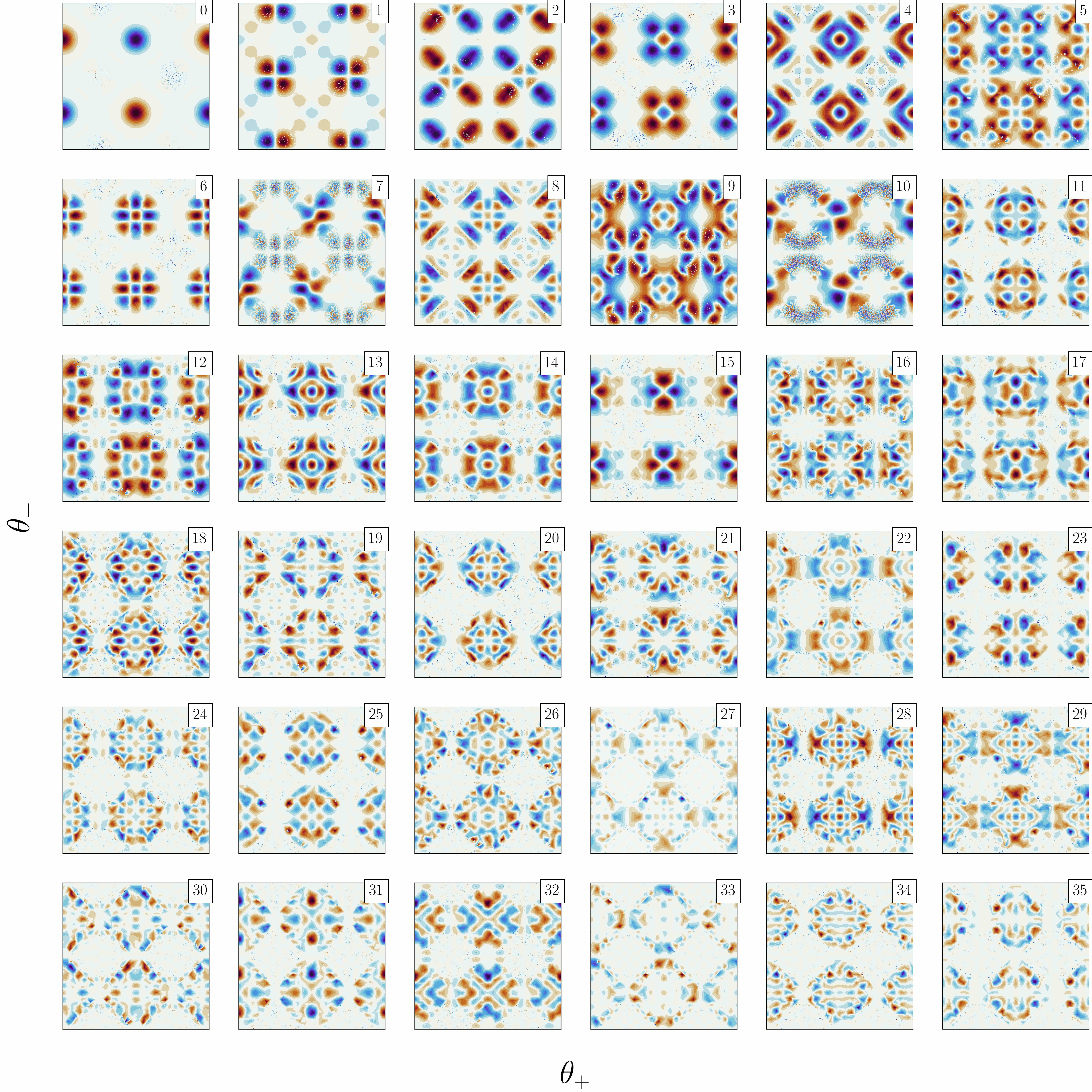}
	\caption{{\bf PCA-transformed invariants antisymmetrized with respect to exchange of atomic sites.}
		Overall, the effect of the PCA is similarly drastic as in the case for the $\vec{\xi}$-symmetric features, leaving us with a handful of relevant features as input.}
	\label{fig:asym_invs_pca}
\end{figure}

When we focus now on the data shown in Figs. \ref{fig:sym_invs_pca} and \ref{fig:asym_invs_pca}, we observe that the PCA transformation succesfully reduces the dimensionality of the feature space, while keeping the characteristics of the different types of invariants intact and separated. Figs. \ref{fig:sym_invs_pca} and \ref{fig:asym_invs_pca} display the features obtained by PCA-transformation from the $\vec{\xi}$-symmetric and $\vec{\xi}$-antisymmetric features as functions of the spherical coordinates $\theta_{\pm}$. The panels show one invariant each, sorted by decreasing explained variance from top left to bottom right. The principal components of the feature space scoring high in explained variance represent the mixtures of types that were identified in the original feature space. Most strikingly however, the polar structures, more associated with higher order invariants, score highest in explained variance. From the data presented in this section, it is apparent that efficient dimensionality reduction can be  achieved by the PCA-transformation, which additionally provides us with invertability (so we can easily transform back to the original space) and conceptual simplicity (since it is a linear transformation).


\subsection{Feature selection with statistical methods}

\begin{figure}[t!]
	\includegraphics[width=.99\hsize]{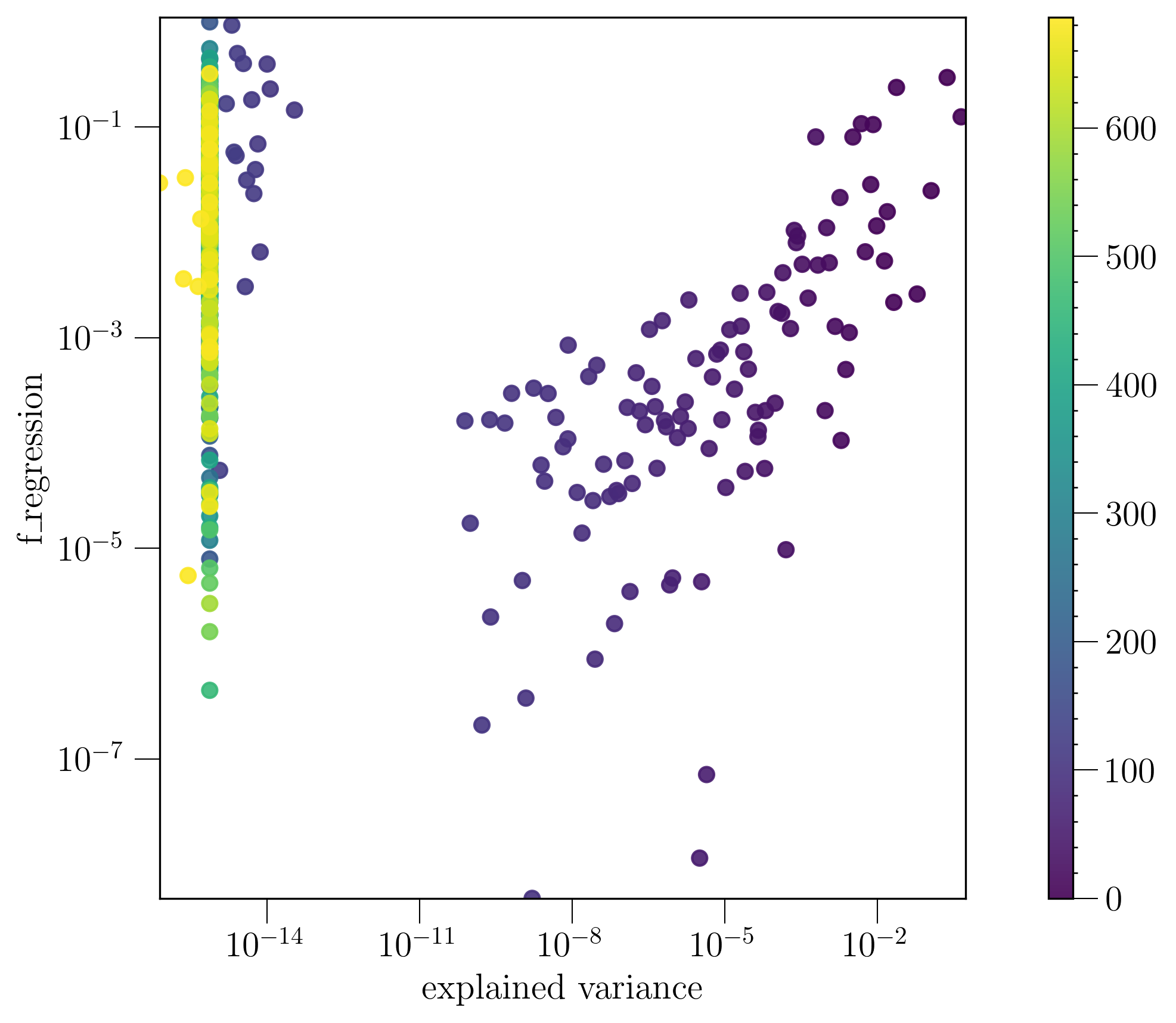}
	\caption{{\bf Explained variance ratio and f-regression score for the $\vec{\xi}$-symmetrized PCA components.} 
		Explained variance ratio of the PCA components on a logarithmic scale on the $x$-axis and f-regression score on a logarithmic scale on the $y$-axis. The PCA transformation finds a number of nontrivial components with nonzero explained variance ratio, corresponding to the true rank of the feature matrix. The sharp step in explained variance around index $110$ indicates the true rank, as the values for the explained variance ratio drop to zero for components beyond this index. Correspondingly, the data fall into two clusters, one with relevant influence on the feature space in the right cluster, and one with negligible influence in the left cluster, correspondingly expressed in the values of explained variance. Notably, the relationship between f-regression score and explained variance shows a linear trend in the right cluster, however the data points show sizable spread. The overall negligible variance and magnitude of components beyond the true rank of the feature matrix, i.e. data points in the left cluster, lead to inflated f-regression scores. We therefor disregard these components scores as uninformative.}
	\label{fig:ExVar_v_FI_sym}
\end{figure}

Investigating the f-regression scores presented in \cref{fig:ExVar_v_FI_sym}, indicating the correlation between single feature and target, we can form a clear picture of which input features are relevant for describing the electric transport. The entries are scattered on the $y$-axis against the values of explained variance on the $x$-axis, in order to relate the component's importance regarding feature space variance and target correlation respectively. On both axes, the data is shown on a logarithmic scale.
Overall, the data is split into two clusters with respect to the explained variance: on the left hand side, redundant components accumulate at the limit of vanishing explained variance. As we are forcing the PCA transformation to construct as many components as there are features, components beyond the true rank of the feature matrix are redundant, having zero explained variance and being constant, as we already established in \cref{sec:PCASection}. Due to the vanishing variance of these components, the f-regression scores in this cluster are highly inflated.

On the right hand side, a cluster of possibly relevant components forms an almost linear trend on the double logarithmic scale. Components below the true rank of the feature matrix are assigned non-vanishing values of explained variance, which show some correlation with the size of the f-regression score. This correlation is a confirmation of the intuitive reasoning, that components with larger portions of explained feature space variance should be more relevant when determining the value of the target on the basis of the features. In contrast to this intuition, the data shows a sizable spread around the trend-line. Accordingly, components associated with higher-order characteristics of the feature space with lower overall variance, still show significant correlation with the target. Measuring the linear correlation of single features and the target is therefore yielding valuable information when performing modelling, which can not be retrieved from feature analysis alone.

\begin{figure*}[t!] \includegraphics[width=0.8\hsize]{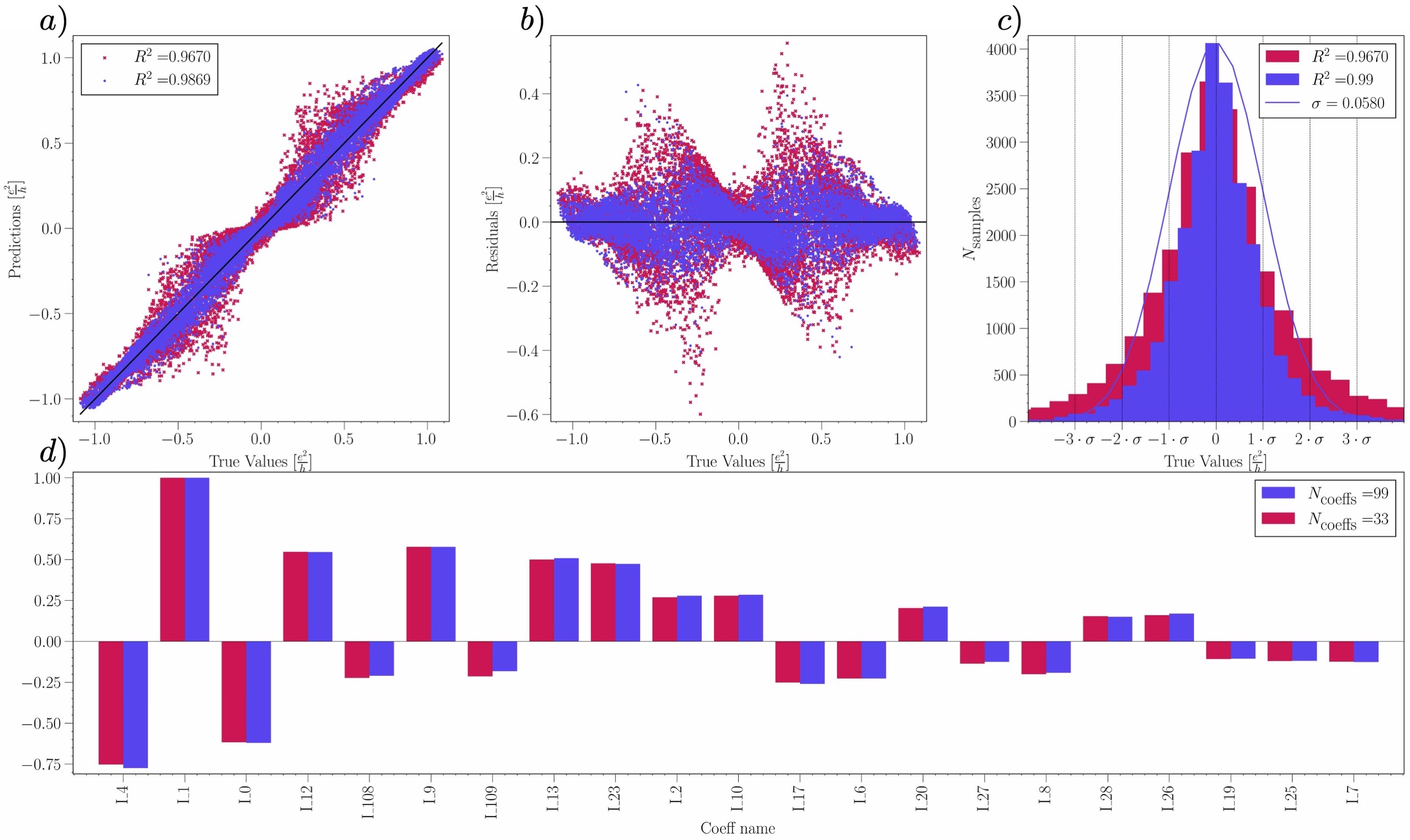}
	\caption{{\bf Fitting results for the anomalous Hall effect.}
		The $\vec{\xi}$-even part of the anomalous Hall conductivity can be well described by a linear model of a few variables. The figure shows correlation plots between true value $y_{\text{true}}$ and fitted target $y_{\text{fit}}$ (a) as well as residual $y_{\text{res}}=y_{\text{true}}-y_{\text{fit}}$ (b), where the red line indicates perfect linear correlation. The distribution of residuals is given in panel (c). Panel (d) presents the coefficients of the model sorted by feature selection score, indicating the linear correlation of one feature and the target estimated by the f-regression test. The model score, shown in the top left corner of panel (a), indicates a strong performance of the model, as the top score that can be achieved is 1. Investigating the coefficients shown in panel (d), it is evident that effective condensation of the input feature space can be achieved by combining PCA, statistical feature selection and model regularization.
	}	\label{fig:FitRes_Trans}
\end{figure*}

\subsection{Model selection: Anomalous Hall effect}\label{sec:FitResults}

Below, we present the results obtained from fitting linear models of the symmetric invariant expansion to the intrinsic part of the anomalous Hall effect (AHE) at a fixed value of Fermi energy $E_{F}=-1.5\;$eV. To this end, for compactness, we focus only on the $\vec{\xi}$-symmetric non-chiral part of the AHE, noting that while the conclusions formulated below generally hold true also for the $\vec{\xi}$-odd part of the AHC, the much smaller average magnitude of the chiral AHC in our system makes the numerical analysis in the chiral case generally more cumbersome. In~\cref{fig:FitRes_Trans} we present the data fitted by two $\vec{\xi}$-even models with different number of PCA-transformed components, selected on the basis of f-regression score, in order to illustrate the importance of using all components up to the true rank of the feature matrix. While the larger model has access to the top $90\%$ of components with respect to the f-regression test (blue crosses), the smaller model only uses the top $30\%$ of components (red dots). This amounts to $99$ and $33$ components out of $110$, respectively.

The first question to be addressed is whether the method used here is able to reproduce the training data with sufficient confidence. In order to rate the model performance we employ the well-known coefficient of determination $R^2$, defined in Eq. \ref{eq:coeffdet}. The best score possible is $1$, and a score of $0$ would correspond to a model simply predicting the mean of the training data for every data point. From the correlation plots presented in Fig.~\ref{fig:FitRes_Trans} (a,b) we can immediately see that both models perform reasonably well, with the large model scoring $0.99$ (blue crosses) and the smaller model $0.97$ (red dots) respectively. 
Obvious deviations from perfect linear correlation (black line) occur in the pockets above and below the point $(0,0)$ in the center of the plot for both models. Such deviations, ranging up to $0.6$ for the model with less coefficients, with the target value being normalized to $1.0$, would be a reason for concern, if they occurred for relevant portions of the test data. However, inspecting the distribution of residuals, shown in Fig.~\ref{fig:FitRes_Trans} (c), proves that only a small number of samples deviate significantly from the center. Inspecting the variance of the residuals for the larger model reveals, that the deviations from the model approximately follow a normal distribution (blue line). From the indicated variance ranges (black dotted lines) it is apparent, that deviations greater than $3\cdot \sigma\approx 0.18$ are quite rare.
By comparing the blue and red distribution, the effect of restricting the model to less components is illustrated by the larger spread of the red distribution. More weight is allocated to the tails of the curve, corresponding to samples with larger difference between target and prediction value.

Turning our attention now to the model coefficients, presented in Fig~\ref{fig:FitRes_Trans} (d), it becomes immediately obvious that the models are sparse. Utilizing the combination of PCA, statistical feature selection and regularization, the method is able to condense the input feature space, containing a total number of $687$ features, to $33$ and $99$ features used in predicting the target value respectively. Allowing the model to use a larger fraction of the input features, as we have seen in discussing the distribution of residuals above, leads to a significant reduction in deviations of predictions from the target. However, the values of the coefficients used in both models do not change drastically, indicating stability of the models. Similarly, when removing small valued coefficients from the fitted models  by hand (not shown here), which we call {\it feature annealing}, has a modest effect on the prediction. Depending on the threshold below which coefficients are annealed, the predictions are relatively stable. This indicates that the annealed coefficients are not relevant to the model and could be suppressed by adjusting the regularization parameter when fitting the model, see cross-validation approach in \cref{sec:Pipeline}.

A special remark can be made for the coefficients of components $108$ and $109$, appearing in \cref{fig:FitRes_Trans} d) in $5^{th}$ and $7^{th}$ place. Keeping in mind that the coefficients are ordered from left to right by f-regression score, these two coefficients are extreme cases of a small explained variance ratio, but comparatively large correlation with the target. Investigating the influence of these two coefficients on the model by hand (not shown here) reveals their strong influence on the model fidelity, in spite of their small relative size, below $25\%$, compared to the largest model coefficient. Annealing these two coefficients from the model increases the maximal residuals by a factor of $1.5$, and $-$ most importantly $-$ the fraction of samples with residuals larger than the $3\sigma$-range is more than doubled. In contrast, removing two coefficients of similar size, but with lower f-regression score, has much less drastic effects on the residuals. In conclusion, f-regression score is a reliable and effective metric for assessing the importance of single components for the model, controlling the overall prediction quality and number of outliers.

Considering that the input features do not contain any information about the electronic properties of the system explicitly, the discrepancy in prediction of the true value might be most probably explained by referring to the band structure, consisting of four bands only~\cite{Kipp2021chiralComPhys}. Tuning the direction of magnetic moments can result in drastic and abrupt changes in the position of the bands with respect to the Fermi energy. By probing the electric transport at constant Fermi energy, the analysis is sensitive to sudden changes in occupation, for example when single bands are pushed above or below the Fermi energy by altering the magnetic configuration. Correspondingly, analysis of the prediction for different energetic positions in the band structure shows, that the amount of deviation in the model prediction depends on the Fermi energy. In order to avoid this nonuniform behavior, the target can be calculated at constant filling. This approach, which we discuss in detail below, requires much more numerical effort, since the calculations of the AHC have to be complemented by precise calculations of the density of states.
In contrast, the expansion could be adapted to incorporate certain band structure effects, sacrificing conceptual simplicity for a more accurate description of the transport. On the other hand, considering the fact that in larger systems the electronic structure contains many more bands, the changes in occupation due to the rearrangement of bands might be less drastic, and the general behavior of the coefficients and fitting much smoother. Correspondingly, deviations in model prediction could be dramatically reduced.

\subsubsection{Model coefficients as function of the Fermi energy}
\begin{figure}[t!]
	\includegraphics[width=0.9\hsize]{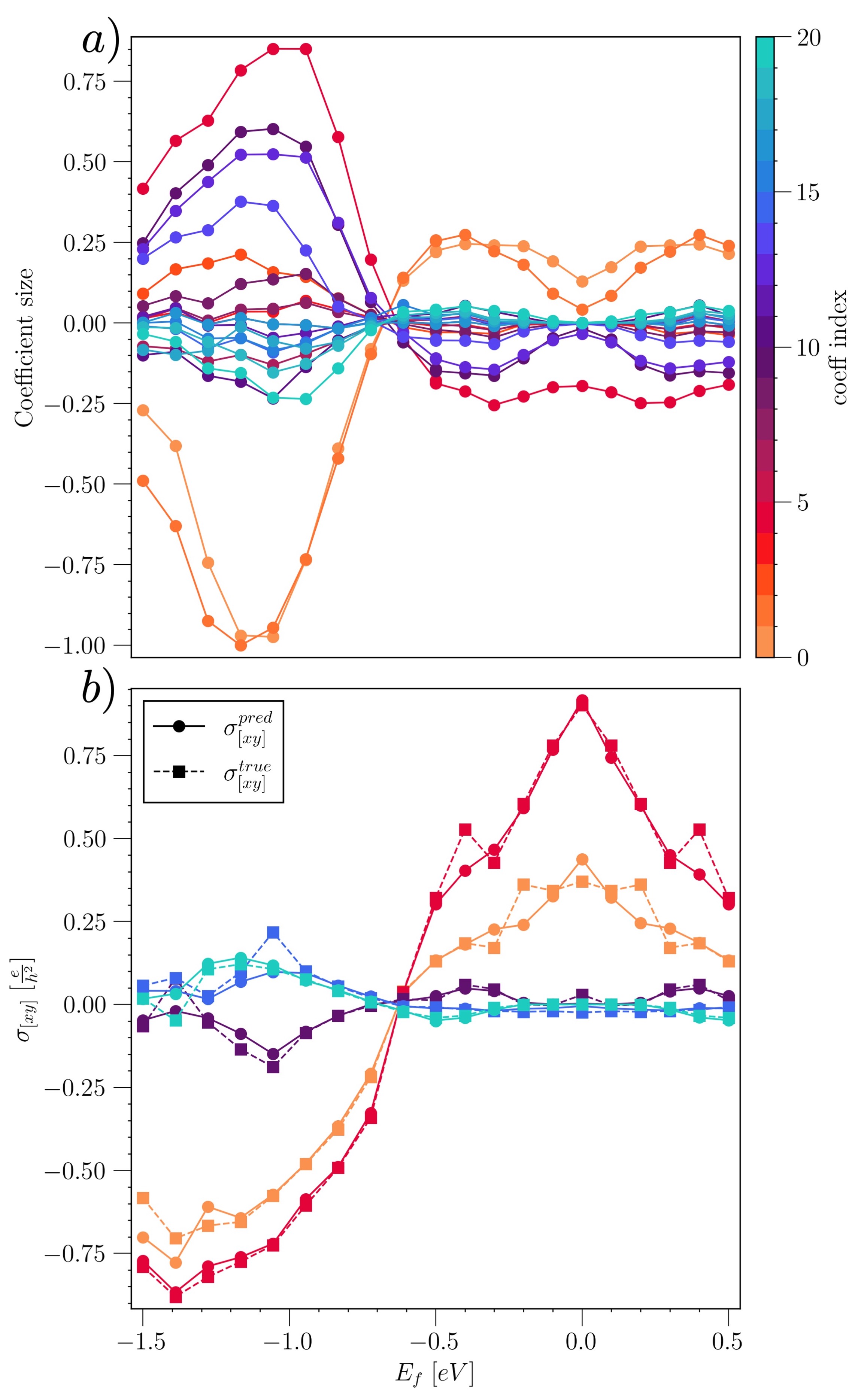}
	\caption{{\bf Fermi energy dependence.} (a) Model coefficients as a function of the Fermi energy. One model is fitted at each of the sampled Fermi energies $E_f$. The model coefficients vary depending on the energy $E_f$ at which the model is fitted. Notably, at $E_f\approx 0.7$ eV, all the coefficients change sign, as does the target. (b)  Predictions of the $\vec{\xi}$-even part of the anomalous Hall conductivity (AHC) as  a function of the Fermi energy. Here, the prediction quality of the model can be inspected for four randomly chosen spin samples (blue, orange, petrol, red), where the true values are shown in dashed lines with square symbols, while the the predictions are shown with solid lines and circles. The data presented here are examples, for which the residuals are below $0.1$.
	}
	\label{fig:coeff_v_fill}
\end{figure}

The AHE is a quantity driven by the occupied states of a system. It is therefore instructive to investigate the influence of the Fermi energy $E_{F}$ on the selected model $-$ we quantify this relationship by fitting models at several values of the Fermi energy and tracking the values of the model coefficients. Only states below $E_{F}$ are occupied and therefore contribute to the AHC, which allows us to associate model coefficients and their size with regimes in the electronic structure contributing to the electric transport. As the data presented in Fig.~\ref{fig:coeff_v_fill}(a) clearly shows, the model coefficients in the $\vec{\xi}$-even case are smooth functions of the energy, which provides a basic sanity check for the model: the selected model transfers from one regime in the electronic structure to another, without selecting a different set of features. We can further identify a few dominating features in the model, with the corresponding model coefficients being as much as five times larger than the other coefficients in size. Perhaps the most significant characteristic of the presented data is the sign change that all coefficients undergo at a Fermi energy of about $ -0.65$ eV, which corresponds to a sign change in the target, or AHE, at that specific energy. Overall, we also observe how the interplay between dominant, lower, and higher-order components of the chiral AHE can sensitively depend on the electronic structure. 

Inspecting the behavior of the predictions and target values for four representative spin configurations (blue, orange, cyan, red) as a function of the Fermi energy, we can immediately recognise that in the even case, Fig.~\ref{fig:coeff_v_fill}(b), the model prediction (circular markers, solid line) resembles the target values (square markers, dashed line) closely across different values of the energy. At the Fermi energy of about $0.7$\,eV, the target and prediction for all samples change sign. The samples we are referring to here are randomly selected from the samples with residuals below $10\%$ at all values of the Fermi energy, which constitutes about $82\%$ of the training set. The examples presented in \cref{fig:coeff_v_fill}(b) are a strong indicator that the model closely represents the true target value on a significant amount of previously unseen test data.

\begin{figure}[t!]
	\includegraphics[width=0.9\hsize]{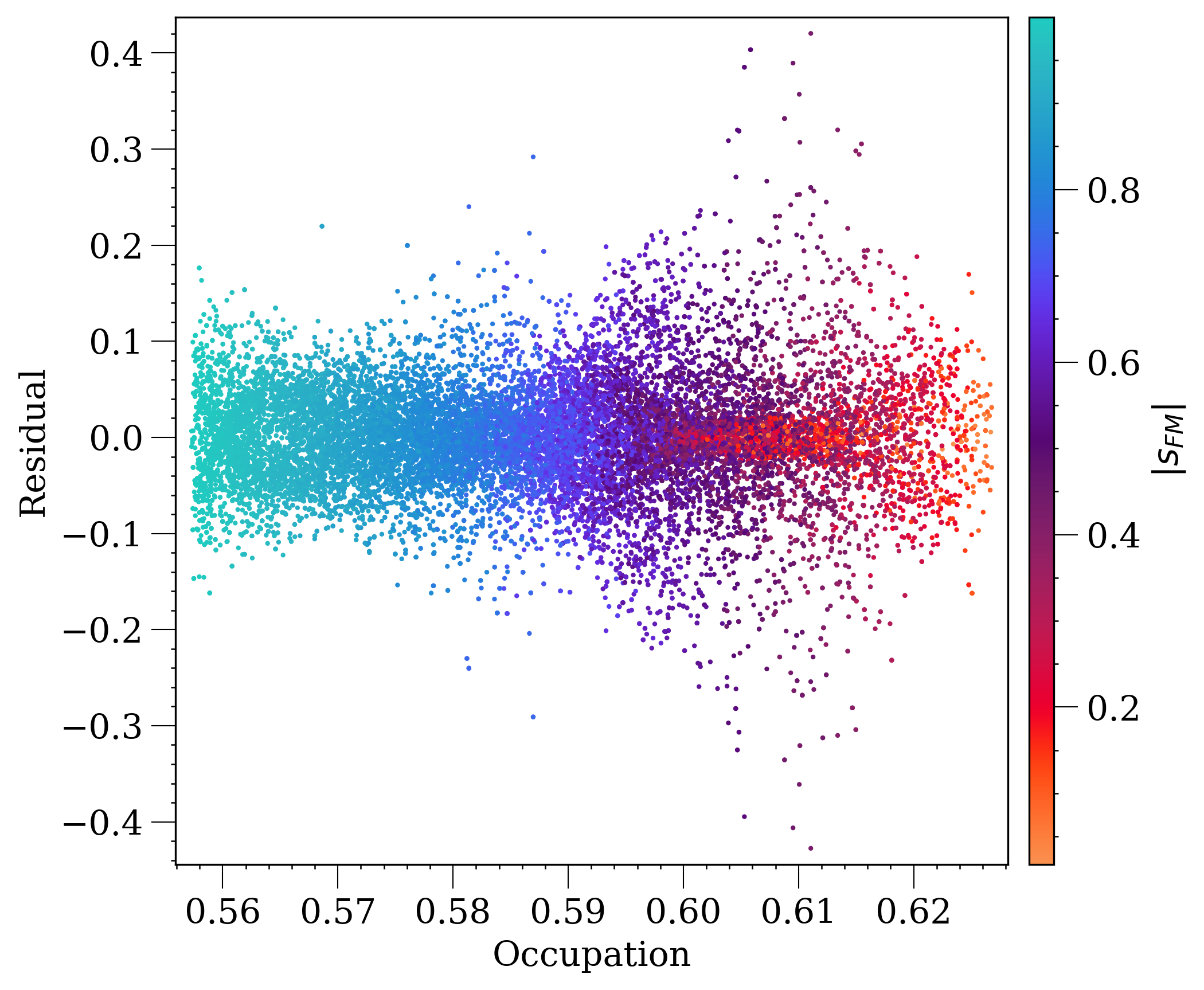}
	\caption{{\bf Residuals of the fitted model as a function of electronic filling.} Deviations of predicted value from true value of the target, $\sigma_{xy,true}-\sigma_{xy,pred}$
		show distinguished pockets below and above zero. Plotting the deviations as a function of electronic filling of the model, which in turn is altered by the magnetic configuration, reveals strong correlation between residuals and filling. At filling value of $\sim0.62$, the model predictions deviate strongly, and the overall linear correlation between filling and absolute value of the ferromagnetic moment breaks.}
	\label{fig:Res_vs_Fill}
\end{figure}

\subsubsection{Transport at constant electronic filling}

Coming back to the systematic deviations of model predictions from the true target, shown in \cref{fig:FitRes_Trans}(b) as pockets below and above zero at the center of the plot, we suggested to identify varying electronic filling as one of the underlying difficulties for a linear model.
Indeed, tuning the magnetic configuration results, as a direct consequence of the presence of spin-orbit interaction in the electronic model \cref{eq:model}, in changes to the energetic position and shape of electronic bands in reciprocal space. As a direct consequence, the electronic occupation at a given value of the Fermi energy is not constant under changes in the magnetic configuration.
In order to judge whether differences in electronic filling, driven by these changes, introduce systematic deviations in the fitted model, we present the residuals of the fitted data as a function of electronic filling, with a color code indicating the absolute value of magnetization, in \cref{fig:Res_vs_Fill}. 

\begin{figure*}[t!]  \includegraphics[width=0.8\hsize]{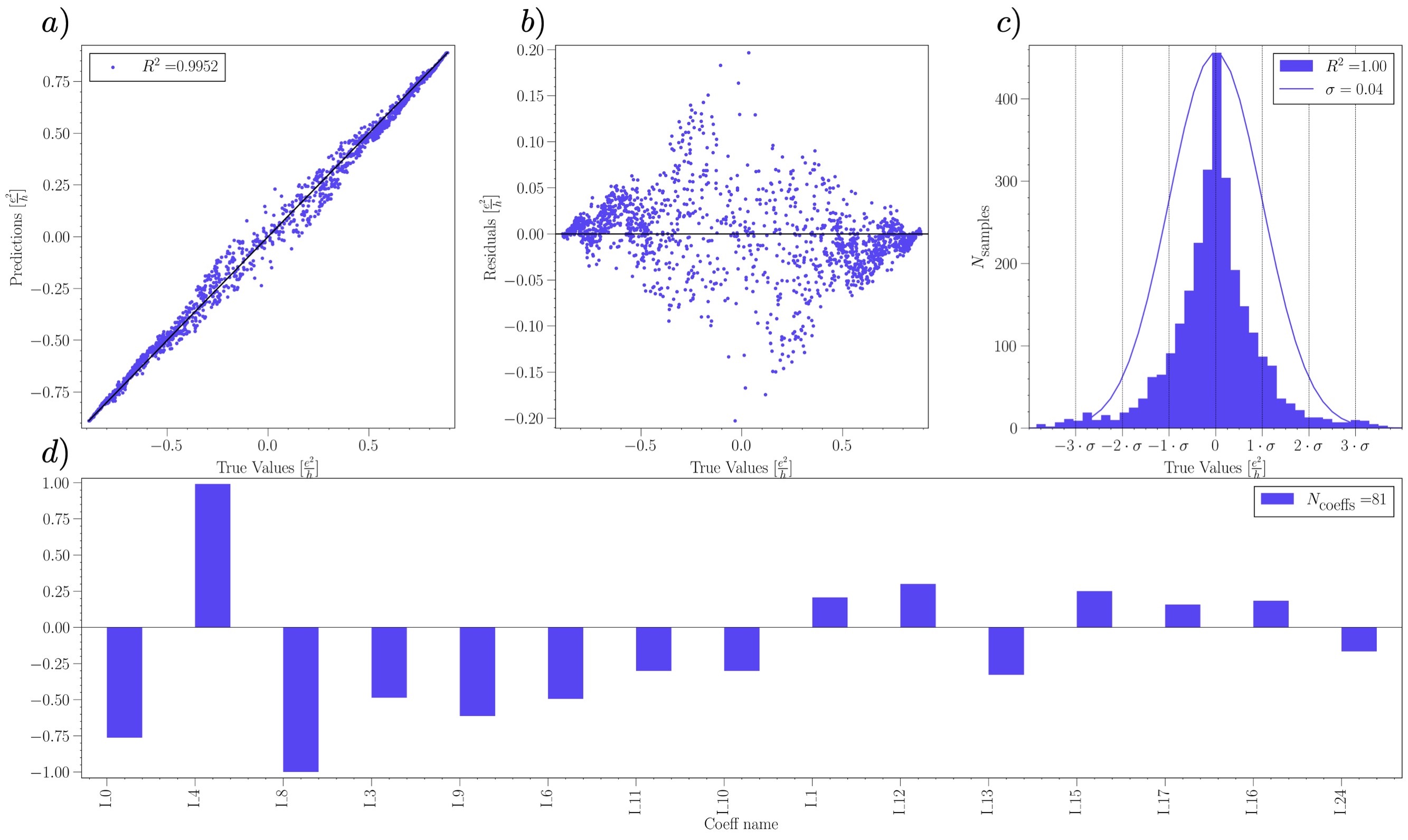}
	\caption{{\bf Fitting results for the anomalous Hall effect at constant electronic filling.} The $\vec{\xi}$-even part of the anomalous Hall effect (AHE) of the two atom system can be well described by a linear model of a few variables also in the case of constant electronic filling. 
		The structure of the plot and corresponding variables are the same as in Fig.~\ref{fig:FitRes_Trans}.
	}
	\label{fig:FitResults_ConstFill}
\end{figure*}

As basic physical intuition concerning  magnetic systems with strong exchange splitting and weak spin-orbit interaction would suggest, the electronic filling of the system, shown on the $x$-axis in the figure, correlates strongly with overall magnetization in the system. For the region of $[0.56,0.6]$ electrons per atom, the residuals are constrained to the range $[-0.1,0.1]$. Most interestingly, the deviations of model prediction from true target value increase suddenly around a value of about $0.61$ for the electronic filling, with a magnitude of up to $0.4$. Furthermore, the strict correlation between occupation and magnetization is lifted at this value, with strongly compensated configuration showing small residuals, and samples with moderate magnetization inhibiting large deviations. We can thus conclude that the residuals of the fitted model show systematic dependence on electronic filling, which is modulated by the magnetic configuration, in a strongly nonlinear fashion. The data presented in \cref{fig:Res_vs_Fill} is strong evidence, supporting the idea that systematic deviations in the linear model are caused by nonlinear characteristics of the electronic structure, induced by changes to the magnetic configuration and mediated by the coupling of spin to electronic degrees of freedom through spin-orbit interaction.

In order to rectify the approach, we present a model fitted to training data calculated at constant electronic filling in \cref{fig:FitResults_ConstFill}. As is the case for \cref{fig:FitRes_Trans}, the model is fitted to $\vec{\xi}$-symmetrized data, with a score larger than $0.99$, and visibly smaller deviations in prediction when comparing the data presented in panels (a), showing model predictions on the $y$-axis and true target value on the $x$-axis, in \cref{fig:FitResults_ConstFill} and \cref{fig:FitRes_Trans}.
Under closer inspection of the data shown in panels (b), which present residuals $y_{\text{true}}-y_{\text{pred}}$ on the $y$-axis and true values on the $x$-axis, we conclude that the overall scale of deviations relative to the absolute value of the target is reduced by factor of about four when considering the effect of constant filling.
The comparison of the model coefficients, presented in panel (d) in each figure, reveals, that very similar components are active in the model, while the precise ordering and assigned weight is slightly different. We can thus conclude that, overall, while accounting for the effect of constant electron filling may somewhat improve the accuracy of the modelling, the qualitative predictions and understanding of the model can be already achieved at the level of a much more computationally efficient  approach which relies on fixing the value of the Fermi energy. The two approaches are expected to converge in the limit of many atoms in the unit cell, as it is the case for example for complex spin textures or fluctuating magnets.

\section{Discussion}\label{sec:Conclusion}
The main goal of this study was to assess whether the electric transport, generated by the intricate changes in electronic structure introduced by complex non-collinear magnetism, can be predicted by using only descriptors of the magnetic pattern realized on a two-dimensional lattice. 
First, in order to obtain a complete, but non-redundant, description of the general conductivity tensor, an expansion in symmetric invariants of the underlying lattice point group was utilized to find a suitable set of descriptors.
Second, a machine-learning pipeline was designed, tested and trained for the purpose of finding a sparse, linear model of the conductivity as a function of the symmetric invariants, which displays reasonable generalization statistics on unseen data. Relevant features of the input data were extracted by utilizing two complementary methods: on the one hand, feature selection was performed by ranking the features with a statistical metric measuring the correlation between feature and target, the f-regression test. On the other hand, regularization of the model search with the elastic net penalty was used to assign lower scores to models using a larger number of features, therefore pushing the search towards models which minimize the number of nonzero coefficients. Combining these two steps has led to significant condensation of the input feature space in conjunction with good overall generalization ability, verified by reasonable scoring of the final model on the test set.

The results of this study illustrate, that an excellent description of electric transport in magnetic materials can be achieved on the basis of the magnetic structure only, when the electronic filling of the system is considered to be unchanged by tuning the magnetic configuration, while the analysis performed for a fixed Fermi energy can already give a very good insight into the qualitative constituents of the model and the relative importance of different terms in the conducticity expansion. Given suitable modelling techniques, explicit calculations of the electronic structure, especially challenging in materials exhibiting strong spin-orbit interaction and hosting complex magnetic textures, can be augmented and enhanced by the symmetric invariants method, encoding the magnetic order parameters of the system.

Commenting especially on the statistical feature analysis, the results suggest the importance of magnetic texture characteristics beyond the established order parameters of ferromagnetic and staggered antiferromagnetic moment. Overall, while the feature space can be condensed considerably, the resulting linear models still show remarkable entanglement of magnetization properties beyond the established order parameters. While the resulting model is linear, the PCA components used as building blocks for the model, are sensitive beyond linear dependence to intricate changes in the orientation of magnetic moments. Our findings thus underline the importance of higher-order in spin contributions for a consistent description of the AHE already in the simplest case of bipartite canted magnets described by elementary electronic models with spin-orbit interactions. This is the main finding of our work concerning the physical properties of quantum spin systems. We can speculate with a high degree of certainly that the situation is even more complex for example in frustrated spin systems of kagome type, where various contributions to 
the AHE are still under debate (see e.g.~\cite{Rajan2023_MnNiCu}). Our approach is thus unique in identifying the magnitude and symmetry of leading AHE contributions without resorting to an excruciating process of guessing complemented by scarce calculations. 

A special word should be said about the potential of suggested methodology in exploring the influence of various further parameters and corresponding more complex phase-spaces on the transport properties of dynamically evolving or fluctuating magnets. Indeed, we expect that the effect of various characteristics of the quantum system as reflected in its Hamiltonian $-$ such as the magnitude and symmetry of the hoppings, strength of correlations, exchange splitting or spin-orbit interaction strength, which can be tuned dynamically~e.g.~by temperature, laser pulse, or other external means $-$ can be consistently included into our analysis. Going beyond, we dare to suggest a possibility of including the time evolution of the spin system according to some dynamical equation into consideration explicitly, which may result in finding a clear path towards  consistent modelling of the memory kernel of the system~\cite{Gouasmi2017_Memory}.

Overall, we can conclude that, in spite of our work focussing on a simple four-band electronic structure component, the principal possibility of modelling that we have demonstrated here marks an important step in integrating machine learning methods into the field of spintronics research and in particular magneto-transport phenomena. Symmetry-enhanced compressive sensing in conjunction with principal component analysis turns out to be a promising candidate to conquer the complexity of magnetic phase-spaces, and the study of spin and electronic transport phenomena offers plenty of avenues for extending the machine learning technique. 
The symmetric invariants, although already showing to be a feature space with impressive potential for generalizability, can be compared and possibly augmented with features directly extracted from the magnetic texture in real space, or the electronic structure in reciprocal space. Specifically, direct feature extraction, e.g. through variational auto encoders (VAE), lends itself to the case of large textures with many atomic sites, when the construction of local chiral and non-chiral contributions is not  trivial,  as opposed to the two-atom system studied here. Acquiring a firm ability to predict the transport properties of large spin textures is pertinent for our ability to model and understand the response characteristics of matter in an automated manner, as such presenting a challenge for research at the junction of solid states physics and machine learning.

\section*{Acknowledgements}
We express our gratitude to Marvin Schmidt, Michael Dick,  Leopoldo Sarra and Florian Marquardt for fruitful discussion and practical guidance on the application of machine learning models to physical data.
Further, we  acknowledge  funding  under SPP 2137 "Skyrmionics" of the DFG.
We  also gratefully acknowledge the J\"ulich Supercomputing Centre and RWTH Aachen University for providing computational resources under project Nos. jiff40 and jpgi11. 
The work was also supported by the Deutsche Forschungsgemeinschaft (DFG, German Research Foundation) $-$ TRR 173 $-$ 268565370 (project A11), TRR 288 – 422213477 (project B06), and project MO 1731/10-1 of the DFG. We also acknowledge funding under the ERC synergy grant 3D Magic, grant number ERC-2019-SyG. 

\section*{Competing interests}

The authors declare no competing interests.


\bibliography{literature}
\end{document}